%% file: 5points3.tex
\documentclass{article}

\usepackage{amsmath}
\usepackage{amssymb}
\usepackage{a4wide}
\usepackage{graphicx}
\usepackage{array}
\usepackage{slashed}
\usepackage{float}
\usepackage{color}
\usepackage[nosort]{cite}
\usepackage{hyperref}

\allowdisplaybreaks[1]

\def\cI{{\cal I}}
\def\cO{{\cal O}}

\topmargin = - 1.5 cm

\begin{document}

\thispagestyle{empty}

\null\vskip-60pt \hfill
\begin{minipage}[t]{4cm}
HU-Mathematik:~2013-23 \\HU-EP-13/75 \\DCPT-13/49\\
\end{minipage}

\vskip1.5truecm
\begin{center}
\vskip 0.2truecm

 {\Large\bf
   Local integrands for the five-point amplitude in planar N=4 SYM up to five loops}
\vskip 1truecm

\vskip 0.7truecm
{\bf    Raquel G. Ambrosio$^{a}$, Burkhard Eden$^{a}$,\\  Timothy Goddard$^{b}$, Paul Heslop$^{b}$,
  Charles Taylor$^{c}$ \\
}

\vskip 0.3truecm
{\it $^{a}$ Institut f\"ur Mathematik und Physik, Humboldt-Universit\"at, \\
Zum gro{\ss}en Windkanal 6,  12489 Berlin
\vskip .2truecm
$^{b}$   Mathematics department, Durham University, Durham DH1 3LE,
UK
\vskip .2truecm
$^{c}$ 8 Cherryl House,
Seymour Gardens,
Sutton Coldfield,
West Midlands,
B74 4ST,
UK
} 
\end{center}

\vskip 2.5 cm 

\centerline{\bf Abstract}
\medskip
\noindent 
{Integrands for colour ordered scattering amplitudes in planar N=4 SYM are dual to those of correlation functions of the energy-momentum multiplet of the theory. The construction can relate amplitudes with different numbers of legs.

By graph theory methods the integrand of the four-point function of energy-momentum multiplets has been constructed  up to six loops in previous work. In this article we extend this analysis to seven loops and use it to construct the full integrand of the five-point amplitude up to five loops, and in the parity even sector to six loops.

All results, both parity even and parity odd, are obtained in a concise local form in dual momentum space and can be displayed efficiently through graphs. We have verified agreement with other local formulae both in terms of supertwistors and scalar momentum integrals as well as BCJ forms where those exist in the literature, i.e. up to three loops.

Finally we note that the four-point correlation function can be extracted directly from the four-point amplitude and so this uncovers a direct link from four- to five-point amplitudes.}

\vspace{0.8cm}

\newpage

\section{Introduction}

There has been a great deal of recent progress in calculating
scattering amplitudes of the maximally
supersymmetric non-abelian Yang-Mills theory in four
dimensions, ${\cal N} = 4$ SYM. In particular, interesting structures
enabling new results have been found
for the amplitude integrand both in the planar
limit as well as for the full non-planar
theory. Perturbative calculations by Feynman graphs are complicated due to the vast number of contributing diagrams, which makes it difficult to construct even the integrand. To evaluate the integrals is, of course, the hardest step --- which we will not attempt to take in this work --- but it will obviously be facilitated by finding simple concise forms of the integrands.

There have been three main methods used for generating integrands.
(Generalised) unitarity is the most widespread
technique~\cite{hep-ph/9403226,hep-ph/9409265,hep-th/0412103,hep-th/0506126}. Here
one equates the leading singularities of an ansatz --- consisting
of a sum of independent graphs with arbitrary coefficients --- with those of the amplitude, which will fix all freedom.
There are various criteria as to which graphs should occur in the
ansatz: in the planar limit one uses dual conformal invariance~\cite{0709.2368,0712.1223,0807.1095,0807.4097,0906.3552},  whereas in the full
non-planar theory one can use the colour-kinematics
duality~\cite{0805.3993,1004.0476}. This technique has been used to
obtain the four-point amplitude up to five-loops
(planar~\cite{hep-ph/9702424,hep-th/0505205,hep-th/0610248,0705.1864}
and nonplanar~\cite{hep-th/9802162,0808.4112,1008.3327,1207.6666}),
the five-point amplitude to
three-loops~\cite{hep-th/0604074,1106.4711,0808.1054} and six-point 
amplitude to two loops~\cite{0803.1465,0805.4832}.

Second, one may employ a recursion relation determining higher loop
amplitudes in terms of lower ones~\cite{nima1}.  The original BCFW
\cite{BCFW} recursion decomposed higher-point {\em tree}-level
amplitudes into products of three-point amplitudes, but  in a striking
development this technique has been generalised to loop
\emph{integrands} \cite{nima1}. The use of these on-shell methods,
in particular in terms of momentum twistor variables, yields
relatively compact expressions where Feynman graph calculations would
often result in millions of terms. By construction, BCFW recursion
leads to non-local integrands, i.e. individual terms have poles which
are not of $1/p^2$ type. Yet, the existence of the Feynman graph
method guarantees the cancellation of such spurious singularities in
the sum of all terms. It remains a formidable problem though, to find
simple local forms for the BCFW output, since the recursion procedure
--- although much more concise than any direct graph calculation ---
does fan out considerably for higher-loop integrands (although much
progress has been made towards a resolution of this problem~\cite{nima3}). So
far explicit formulae for local integrands via this method are available for MHV $n$-point amplitudes up to three loops and NMHV $n$-point amplitudes up to two loops \cite{nima1,nima2}.

Third, another less widely known but extremely powerful technique
starts from an ansatz, but now fixes the coefficients by implementing the
exponentiation of infrared singularities at the level of the integrand
by asserting that the log of the amplitude should have a reduced
singularity~\cite{1112.6432}. This method has been used to obtain the
four-point amplitude to seven loops~\cite{1112.6432}, and has been shown
to determine the $n$-point amplitudes at two and three loops for any $n$~\cite{1203.1915}.

Both this method and generalised unitarity customarily use graphs
with local integrands. In addition, the trial graphs used in generalised
unitarity methods typically contain only Lorentz products, with any parity odd structures being in the external variables only. 

Planar scattering amplitudes in ${\cal N}=4$ are dual to polygonal
Wilson loops with light-like edges~\cite{0705.0303,0707.0243,0707.1153,0803.1465,0803.1466}. It has recently been
shown that both sides of this duality can be generated from $n$-point
functions of the energy-momentum tensor multiplet of the theory
\cite{1007.3246,1007.3243,1009.2488,1103.3714,1103.4353}\footnote{Other operators are also suited, see
  \cite{1103.4119}.} To this end, the operators in such an
$n$-point correlator are put on the vertices of an $n$-gon with
light-like edges. The relation between correlation functions and
Wilson loops, which are also
defined on configuration space, is rather
direct~\cite{1007.3243} and can be made
supersymmetric~\cite{1009.2225,1010.1167,1103.4119}.
On the other hand, the connection between energy-momentum correlation functions and amplitudes is conceptually not well understood, while it provides a fully
supersymmetric integrand duality which exactly reproduces the BCFW based loop integrands \cite{1009.2488,1103.3714,1103.4353,1103.4119}. The counterpart of the disc planarity of amplitudes is planarity on the sphere for the correlation functions. More specifically, the correlation functions yield the square of the amplitude integrands; here the two discs are quite literally welded together like the hemispheres of a ball touching at the equator.

The component operators in the energy momentum tensor multiplet are dual  to  supergravity states on AdS$_5$ in the AdS/CFT correspondence \cite{123}. In the next section we will use superspace to package all the component operators into one superoperator; he we rather write $\cO_{\Lambda}(x)$ where $\Lambda$ is simply a schematic label describing the precise component in question.

Two- and three-point functions of these
operators can be shown to be protected from quantum corrections. As
the first non-trivial objects the four-point functions have been intensely studied both at weak coupling
in field theory perturbation theory and at strong coupling exploiting
the AdS/CFT duality. The loop corrections to these four-point function
take a factorised form \cite{partial}:
\begin{equation}\label{eq:1}
 \langle {\cal O}_{\Lambda_1} {\cal O}_{\Lambda_2} {\cal O}_{\Lambda_3} {\cal O}_{\Lambda_4} \rangle
\, = \,  \langle {\cal O}_{\Lambda_1} {\cal O}_{\Lambda_2} {\cal O}_{\Lambda_3} {\cal O}_{\Lambda_4}
\rangle_{\mathrm{tree}} + {\cal I}_{{\Lambda}_1 {\Lambda}_2 {\Lambda}_3 {\Lambda}_4}(x_i) \times \, f(x_i;a)
\end{equation}
In this equation ${\cal I}$ does not depend on the 't Hooft coupling
$a = g^2 N/(4 \pi^2)$ but does depend on the particular component
operators in question; whereas all the non-trivial coupling dependence
lies in the single function $f$.
We have $\ell$-loop  integrands  $f^{(\ell)}(x_1, \dots x_{4+\ell})$ via\footnote{Note that we use the same symbol here $f^{(\ell)}$ and throughout for the integrated function as well as the integrand.}  
 \begin{align}
   \label{eq:26}
   f(x_i;a)&=\sum_{\ell=1}^\infty  {a^\ell\over \ell!} \,\int d^4x_5 \dots d^4x_{4+\ell} \;
   f^{(\ell)}(x_1, \dots, x_{4+\ell})
 \end{align}

The one- and two-loop contributions were computed using supergraphs
\cite{EHSSW,ESS}. In~\cite{hidden,constructing} it was shown that all
the loop integrands  have  an unexpected (hidden) symmetry permuting internal and external variables:
\begin{align}
  \label{eq:27}
  f^{(\ell)}(x_1, \dots x_{4+\ell})= f^{(\ell)}(x_{\sigma_1},\dots x_{\sigma_{4+\ell}})   \qquad
  \qquad \forall \, \sigma \in S_{4+\ell}\ .
\end{align}
The $S_{4+\ell}$ invariance together with conformal covariance
($f^{(\ell)}$ must have conformal weight 4 at each of the $4+\ell$
points),  the absence of double propagator terms (which follows from an
OPE analysis), and planarity of the
corresponding graph beyond 1 loop, constrains the number of undetermined parameters
in an ansatz of this type so severely that up to three loops there is
only one term in the ansatz. Indeed even at higher loops it was possible to
determine $f^{(l)}$ up to $l = 6$ in
combination with the aforementioned criteria about the exponentiation of infrared singularities
\cite{hidden,constructing}.

We note that any single term in $f^{(\ell)}$ has numerator and
denominator composed of squared distances $x_{ij}^2$. The graph
obtained by regarding the denominator factors as edges is called an
``$f$-graph" below. These provide  an exceptionally compact way to
display the result, for example we display the full four-point
correlator up to five-loops compactly via $f$-graphs below in~(\ref{eq:6}). In diagrams we denote the numerator factors by dashed lines.

At four-points the amplitude/correlation function duality
relates the four-point light-like limit of $f(x_i;a)$ to the four-point amplitude $M_4(x_i;a)$ (divided by the tree
amplitude) in dual momentum space $p_i \, = \, x_i-x_{i+1} $ :
\begin{align}
  \label{eq:28}
  1+ 2\sum_{\ell\geq 0} a^\ell \,F_4^{(\ell)} \  &=\   \bigl( M_4(x_i;a)
  \bigr)^2
\end{align}
where
\begin{align}
F_4^{(\ell)}(x_1,x_2,x_3,x_4)\ &=\  \text{external factor } \times \lim_{\substack{x_{i \, i+1}^2    \rightarrow 0 \\ (\text{mod } 4)}} 
   \int d^4x_5 \dots d^4x_{4+\ell} 
  \,{f^{(\ell)}\over \ell!} \ .
\end{align}
and where  the external factor is simply $ {x^2_{13}x^2_{24}}
\prod_{1\leq i<j\leq 4}  x^2_{ij}$.
Graphically the light-like limit on the l.h.s. corresponds to selecting
all possible 4-cycles in the $f$-graph (corresponding to the four external
points) which then splits the planar f-graph into two disc planar
pieces corresponding to the product of two amplitudes.

The interaction between four-point correlation functions and
amplitudes has been the focus of much work in this
direction~\cite{hidden,constructing,1202.5733}. Indeed one can use this relation in reverse to read off the correlation
function from the amplitude and to this end $f^{(7)}$ has recently been
obtained~\cite{charles} using the corresponding seven-loop amplitude~\cite{1112.6432}.

However, less use has been made of the fact that the very same
four-point correlation function is related to particular combinations
of {\em higher} point amplitudes. This remarkable feature takes place simply
due to the fact that loop corrections of correlation functions are
correlation functions with the Lagrangian inserted. But the
Lagrangian is itself an operator in the energy-momentum supermultiplet
and therefore we find that loop corrections of $n$-point correlators of energy-momentum
multiplets are given by certain  higher point correlators  of
energy-momentum multiplets. These are then in turn related to higher
point amplitudes via the amplitude/correlation functions duality. The
details of how this works will be derived in the next section, but
here let us simply note the result
\begin{align}
  \sum_{\ell \geq 0} {a^\ell } {F_5^{(\ell)}}  \ = \ M_5 \, \overline{M}_5\ ,\label{eq:10}
\end{align}
where $F_5^{(\ell)}(x_1,\dots ,x_5)$ is constructed from the four-point
correlator integrands $f^{(\ell)}$:
\begin{align}
F_5^{(\ell)}(x_1,\dots ,x_5):=   \text{external factor} \times \lim_{\substack{x_{i \, i+1}^2
    \rightarrow 0 \\ (i=1 \dots 5)}}  \int d^4x_6 \dots d^4x_{5+\ell}
 {} {f^{(\ell+1)} \over  \ell! }\ ,
\end{align}
where here the external factor is $1/f^{(1)}=\prod_{1\leq i<j\leq 5}x^2_{ij}$.
$M_5$ is the 5-point MHV amplitude (divided by tree-level)
One can readily see the similaity between~(\ref{eq:10}) and the
four-point relation~(\ref{eq:28}). 
So the {\em{five}}-point light-like limit of the {\em four}-point correlator
$f$-graphs
yields the above combination of the five-point amplitude,
whereas the {\em{four}}-point lightlike limit of the same correlator yields the four-point
amplitude: both the four- and five-point amplitudes are contained in
the four-point correlator!

Even so, how can the single equation (\ref{eq:10}) uniquely determine $M_5$? The perturbative expansion of the r.h.s. contains the parity even part $M_5 + \overline{M}_5$ (by choosing the leading 1 in either factor) but beyond it also all possible product terms.
Now, the (sphere) planar part of the correlator integrand on the
l.h.s. of the equation breaks into classes of terms in exactly the
same way. Taking the five-point light-like limit corresponds to chosing a
5-cycle on the $f$-graph (as opposed to a 4-cycle when considering the
four-point amplitude) which splits the $f$-graph into two disc planar pieces; the $\ell$-loop integrand contains terms corresponding to a single $\ell$-loop integral as well as products of $m$-loop and $(\ell-m)$-loop integrals. The single equation (\ref{eq:10}) is therefore ``stratified'' into an over-determined system that turns out to be beautifully consistent.

The article is organised as follows: In Section 2 we demonstrate how the step from four-point to five-point integrands is taken. The resulting equation is split into classes of products. As a first application we discuss why four-point graphs always appear in a symmetric sum over the position of their massive leg. Sections
3,4,5, discuss the one-, two- and higher-loop amplitudes. Our main
result --- a local form of the complete four-loop amplitude --- is
given in Section 6. Furthermore, with the publication we include
computer readable files containing also the complete five-loop and the
parity even sector of the six-loop integrand in a local form. In a
final section of the actual text we discuss the relation to other
 forms of the amplitude where available in the literature. Some appendices discuss technical details.

\section{The amplitude${}_5$/correlator${}_4$ duality}

\subsection{Deriving the duality}

We here derive and give more detail to some of the main formulae of the
introduction. The starting point is the correlator/amplitude
duality~\cite{1007.3246,1007.3243,1009.2488,1103.3714,1103.4119,1103.4353}. To make the full duality precise we use
superspace to package together component fields.
The components of the energy-momentum tensor multiplet, denoted
$\cO_\Lambda(x)$ in the introduction,
can all be assembled into a single superfield
${\cal O}(x,\rho,\bar \rho,y) = \mathrm{Tr}(W^2)$ where the trace is
over the $SU(N)$ gauge group, see \cite{Howe:1995md,hep-th/9412147} and references therein. The field strength multiplet $W(x,\rho,\bar \rho,y)$ lives on analytic superspace, which combines the Minkowski space variable $x$ with Grassmann odd coordinates $\rho, \bar \rho$ and $y$ coordinates which parametrise the internal symmetry of the ${\cal N}=4$ model.\footnote{Analytic superspace was first introduced for a superspace description of the ${\cal N}=2$ matter multiplet \cite{harmonicBook}.} Likewise, amplitudes connected by supersymmetry can be packaged into a superamplitude customarily parametrised by momentum supertwistors \cite{hodges,0909.0250}.

To obtain the full  duality between any amplitude and any correlation
function of these operators,  one identifies light-like coordinate differences on the correlator side with the ingoing momenta of the amplitude according to $x_{i \, i+1} = p_i$ and puts $\bar \rho$ to zero at all points. The precise identification of the left handed Grassmann odd coordinates $\{\rho_i\}$ with the odd part of the momentum supertwistors $\{\chi_i\}$ is known \cite{1103.3714,1103.4353}, but it is not needed here. Also, in the amplitude limits the $y$ coordinates will factor out.

Now let ${\cal G}_n$ denote the $n$-point function of energy-momentum multiplets ${\cal O}$. The amplitude/correlator duality \cite{1007.3246,1007.3243,1009.2488,1103.3714,1103.4353} states that
\begin{equation}
\lim_{x_{i \, i+1}^2 \rightarrow 0} \frac{{\cal G}_n}{{\cal G}_n^\mathrm{tree}} \, = \, ({\cal M}_n)^2 \, , \qquad \bar \rho = 0 \, . \label{dual}
\end{equation}
On the left hand side of this equation ${\cal G}_n$ is a superspace object containing component $n$-point correlators of any operator in the energy-momentum multiplet in one object (some of which are eliminated by sending $\bar \rho$ to zero);  similarly on the right-hand side ${\cal M}_n$ contains all $n$-point amplitudes in the theory packaged in one superspace object: the superamplitude. To be precise, the symbol ${\cal M}_n$ in the last equation denotes the full superamplitude divided by the tree-level MHV amplitude, so the leading term is 1. Both sides of the equation have expansions both in powers of the odd superspace variables as well as in the coupling constant. Expanding in odd superspace variables we write
\begin{equation}
{\cal G}_n \, = \, \sum_{k=0}^{n-4} G_{n;k} \, , \qquad {\cal M}_n \, = \, \sum_{k=0}^{n-4} M_{n;k}
\end{equation}
where $G_{n;k}$ and $M_{n;k}$ contain $4 k$ powers of the odd superspace variable. In particular, ${ M}_{n;k}$ is the N$^k$MHV superamplitude.

By differentiation in the coupling constant it can be shown that
\begin{equation}
G_{n,k}^{(\ell)} \, = \, \frac{a^\ell}{\ell!} \prod_{i=1}^\ell \left( \int d^4x_{n+i} \, d^4\rho_{n+i} \right) G_{n+\ell;k+\ell}^{(0)} \, , \qquad \ell > 0 \label{loops}
\end{equation}
where the superscript indicates the loop order. In other words, the
$\ell$-loop correction to an energy-momentum  $n$-point function is
given by a superspace integral over a Born level correlator of the
same type, just with correspondingly more points. This opens the
possibility of considering various $n$-gon limits of the same
correlator. We currently know very little about the correlation functions
$G_{n;k}$ with $k<n-4$. On the other hand
following~\cite{hidden,constructing} we have a wealth of information
about the ``maximally nilpotent'' case $k=n-4$.
In this paper we exploit this mechanism to construct the five-point
amplitude from the correlators $G^{(0)}_{n;n-4}$ that were originally elaborated for the higher-loop integrands of the four-point function. Specialising (\ref{loops}) to this case:
\begin{equation}
G_{4,0}^{(\ell)} \, = \, \frac{a^\ell}{\ell!} \prod_{i=1}^\ell \left( \int d^4x_{4+i} \, d^4\rho_{4+i} \right) G_{4+\ell;\ell}^{(0)} \, , \qquad G_{5,1}^{(\ell-1)} \, = \, \frac{a^{\ell-1}}{(\ell-1)!} \prod_{i=1}^{\ell-1} \left( \int d^4x_{5+i} \, d^4\rho_{5+i} \right) G_{4+\ell;\ell}^{(0)} \, . \label{loops45}
\end{equation}

According to \cite{partial,EHSSW,ESS,hidden,constructing} the Born level
correlator with maximum $k=n-4$ (maximally nilpotent piece) has the form
\begin{equation}
G^{(0)}_{4+\ell;\ell}|_{\rho_5^4 \dots \rho_{4+\ell}^4}
 \, =  \, {\cal I}_{1234} \ \rho_5^4 \ldots \rho_{4 + \ell}^4 \ f^{(l)}(x_1, \ldots, x_{4 +\ell}) \, ,
\label{partialNonRen}
\end{equation}
where
\begin{equation}
{\cal I}_{1234} \, = \,  \frac{2 \, (N^2 - 1)}{(4 \pi^2)^4} \, (x_{12}^2 x_{13}^2 x_{14}^2 x_{23}^2 x_{24}^2 x_{34}^2) \, \left( \frac{y_{12}^2}{x_{12}^2}  \frac{y_{23}^2}{x_{23}^2} \frac{y_{34}^2}{x_{34}^2} \frac{y_{14}^2}{x_{14}^2} \ x_{13}^2 x_{24}^2 + \, \ldots \right) \,  \label{defR}
\end{equation}
Here the dots indicate terms subleading in both the 4-gon $x_{12}^2,
x_{23}^2, x_{34}^2, x_{41}^2 \, \rightarrow \, 0$ and the 5-gon limit $x_{12}^2,
x_{23}^2, x_{34}^2, x_{45}^2, x_{51}^2 \, \rightarrow \, 0$ which we
are interested in.

The objects  $f^{(\ell)}(x_1, \dots
x_{4+\ell})$, as explained in the introduction,
are rational, symmetric in all $4+\ell$ variables,
conformally covariant with weight 4 at each point and have no double
poles. They can be displayed graphically via so-called $f$-graphs with
vertices $x_i$ and edges denoting propagators $1/x_{ij}^2$.
From 2-loops in the planar theory, the $f$-graphs will be planar (if
we exclude numerator edges)
$4+\ell$-point graphs with vertices of degree (or valency) four or
more. Since we sum over all permutations of the vertices we need not
label the graph - we sum over all possible labellings. Any vertex with
degree $d$ greater that 4 must be accompanied by $d-4$ numerator lines
to bring the total number of numerator lines  minus denominator lines
equal to 4 (corresponding to the fact that the $f^{(\ell)}$ has
conformal weight 4 at each (external and internal) point) although we
sometimes suppress the numerator lines for visual simplicity.

For illustration we here give the $f$-graphs to five-loops (ie the
four-point correlator up to five-loops) and corresponding expressions up to
three-loops: 

\qquad   \begin{tabular}{rm{1.5cm}m{5cm}}
    $f^{(1)}$ =& \includegraphics[height=1.5cm]{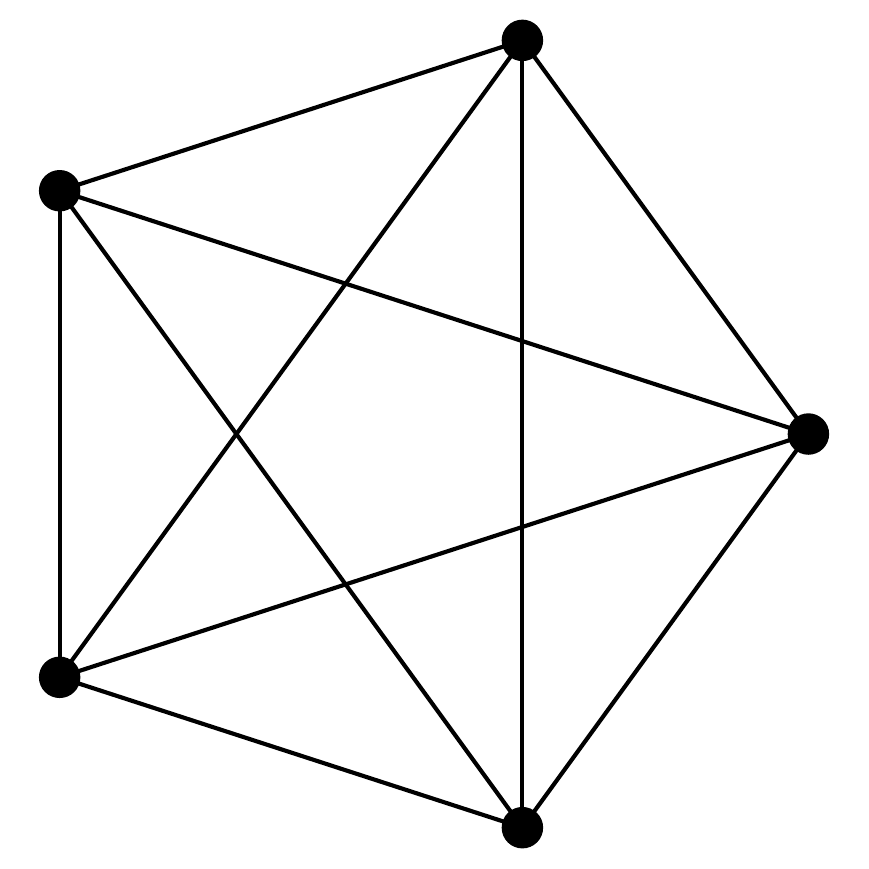}&$ =  \frac{1}{\prod_{1 \leq i < j \leq 5} x_{ij}^2} \, , $\\
    $f^{(2)}$ =&  
    \includegraphics[height=1.5cm]{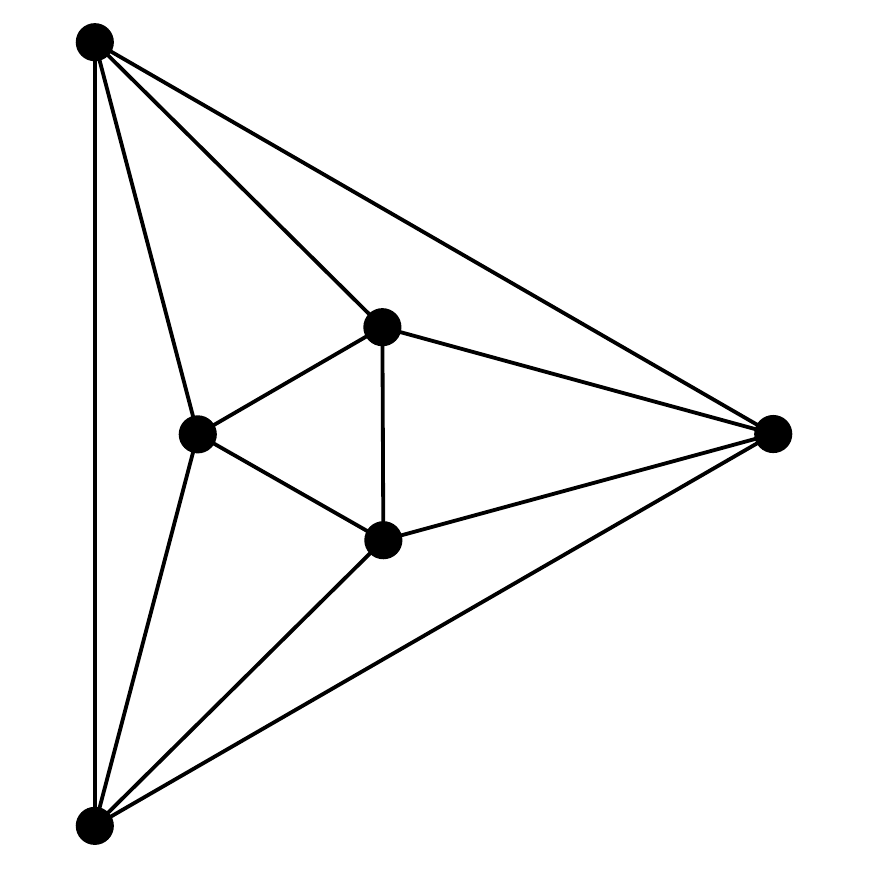} &$=  \frac{\frac{1}{48} \sum_{\sigma \in S_6} x_{\sigma_1
        \sigma_2}^2 x_{\sigma_3 \sigma_4}^2 x_{\sigma_5
        \sigma_6}^2}{\prod_{1 \leq i < j \leq 6} x_{ij}^2}$ 
\\
    $f^{(3)}$ =&\includegraphics[height=1.5cm]{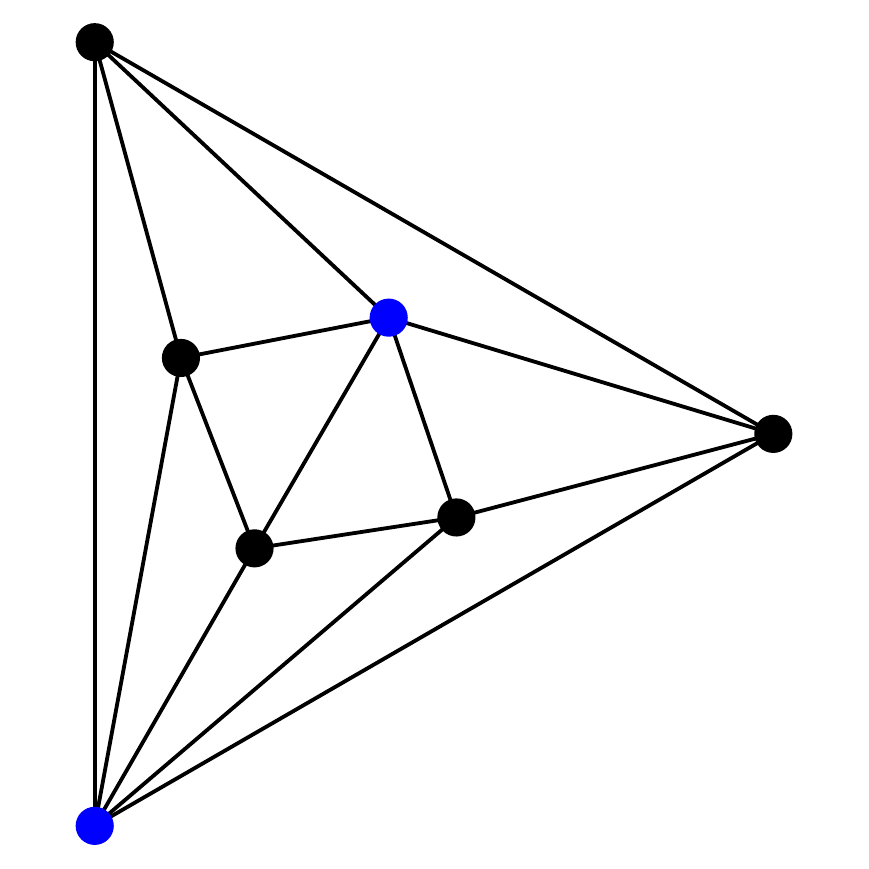}& $=  {{1 \over 20} \sum_{\sigma \in S_7} x_{\sigma_1\sigma_2}^4 x_{\sigma_3\sigma_4}^2 x_{\sigma_4\sigma_5}^2 x_{\sigma_5\sigma_6}^2 x_{\sigma_6\sigma_7}^2
   x_{\sigma_7\sigma_3}^2 \over \prod_{1\leq i<j \leq 7}
    x_{ij}^2}$
\end{tabular}

\qquad   \begin{tabular}{rm{10cm}}
$f^{(4)}=$&\includegraphics[height=1.5cm]{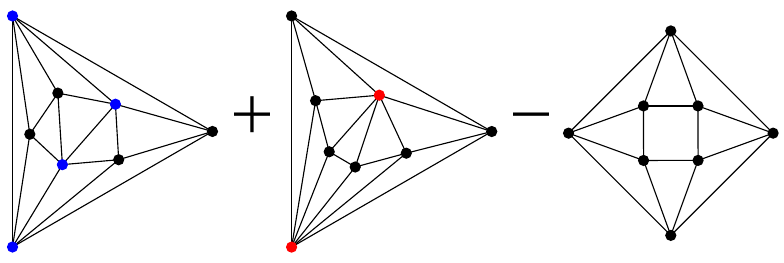}\\
$f^{(5)}=$&\includegraphics[height=1.5cm]{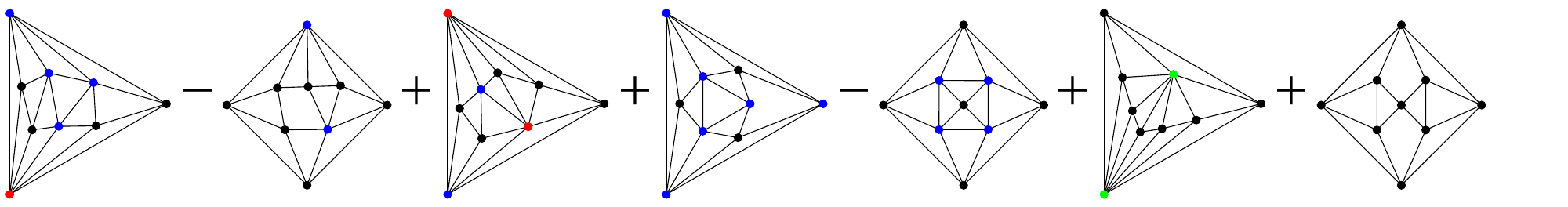}
\end{tabular}

\vspace{-5cm}

\begin{align}
  \label{eq:6}
  \end{align}
 
\vspace{4cm}

We see that $f^{(2)}$ has no remaining numerator terms (all three
apparent numerator terms will be
cancelled by the denominator) whereas $f^{(3)}$ has a single
numerator line (coming from the $x^4_{\sigma_1 \sigma_2}$ in the
numerator which is only partially cancelled by the denominator.) This
numerator edge will connect the two 5-valent vertices (shown in blue).

The one- and two-loop contributions were originally computed using supergraphs
\cite{EHSSW,ESS} whereas the three-loop and higher were computed using
the 
above symmetry considerations (as well as
suppression of singularities for the
coefficients)~\cite{hidden,constructing}.

Now according to~(\ref{loops}), (\ref{loops45}) we can consider this as either a
four-point $\ell$-loop correlator or a five-point $\ell-1$ loop
correlator (or of course a higher point correlator). First let us consider the four-point case (which is the one
focussed on in previous work). 

\subsubsection*{Four-point case}

Eqns~(\ref{loops})
and~(\ref{partialNonRen}) lead directly to the factorised form
\begin{equation}
{\cal G}_4|_{\bar \rho_i = 0} \, =  \, \langle {\cal O}_1 {\cal O}_2 {\cal O}_3 {\cal O}_4 \rangle_{\bar \rho_i = 0}
\, = \, {\cal G}^\mathrm{tree}_4|_{\bar \rho_i = 0} + {\cal I}_{1234}
(x_i, \rho_i,y_i) \, f(x_i;a)
\end{equation}
which is just the superspace version of the factorisation
mentioned in~(\ref{eq:1}).

Now the four-point amplitude/correlator duality~(\ref{dual}) gives
the amplitude purely in terms of $f(x_i;a)$ which we displayed in the
introduction~(\ref{eq:28})
\begin{align}
  \label{eq:28b}
  1+ 2 \sum_{\ell\geq 0} a^\ell \, F_4^{(\ell)} \  &=\   \bigl( M_{4;0}(x_i;a)
  \bigr)^2
\end{align}
where
\begin{align}
F_4^{(\ell)}\ &=\  \text{external factor } \times \lim_{\substack{x_{i \, i+1}^2    \rightarrow 0 \\ (\text{mod } 4)}} 
   \int d^4x_5 \dots d^4x_{4+\ell} 
  \,{f^{(\ell)}\over \ell!} \ .
\end{align}

\subsubsection*{Five-point case}

Let us instead now consider~(\ref{partialNonRen}) as the $\rho_5^4$
component of a five-point correlation function.   For this special
choice equation (\ref{loops}, \ref{loops45}, \ref{partialNonRen}) can be written
\begin{equation}\label{eq:2}
G_{5;1}^{(l)}|_{\rho_5^4} \, = \, \frac{a^\ell}{\ell!}
\prod_{i=6}^{5+\ell} \left( \int d^4x_i \,  \right)G^{(0)}_{5;1}|_{\rho_5^4}  \, \frac{f^{(l+1)}(x_1,\ldots,x_{5+l})}{f^{(1)}(x_1,\ldots,x_5)} \, .
\end{equation}

Now at five points there are MHV and NMHV amplitudes only and NMHV
amplitudes are $\overline{\text{MHV}}$ amplitudes. Therefore
\begin{equation}
M_{5;1} \, =  \, R_{12345} \, \overline{M}_{5;0} 
\end{equation}
where $R_{12345}$ is the five-point $R$ invariant
\cite{0807.1095,0909.0250}. Since there is only one independent object we will henceforth drop the second subscript on $M_{5;0}$ and write $M_5$ instead. Furthermore, in the pentagon light-cone limit
\begin{equation}\label{eq:3}
\lim_{x_{i \, i+1}^2 \rightarrow 0} \frac{G_{5;1}^{(0)}}{G_{5;0}^\mathrm{tree}}\, = \, 2 \, R_{12345}
\end{equation}
as has been shown in \cite{1103.4353}.
The correlator amplitude duality (\ref{dual}) then implies
\begin{equation}
\lim_{x_{i \, i + 1}^2 \rightarrow 0} \frac{G_{5;1}}{G_5^\mathrm{tree}} \, = \, 2 \, R_{12345} \, M_5 \, \overline{M}_5 \, . \label{master1}
\end{equation}
So combining~(\ref{eq:2},~\ref{master1},~\ref{eq:3}) and dividing by
$2 \, R_{12345}|_{\rho_5^4}$  we obtain directly the relation between
$f(x_i;a)$ and the five-point amplitudes quoted in the introduction
\begin{align}
\sum_{\ell \geq 0} {a^\ell } {F_5^{(\ell)}}  \ = \ M_5 \, \overline{M}_5\  \label{eq:10b}
\end{align}
with 
\begin{align}
F_5^{(\ell)}:=  \lim_{\substack{x_{i \, i+1}^2 \rightarrow 0 \\
    (\text{mod } 5)}} 
 {} {f^{(\ell+1)} \over  \ell! \, f^{(1)}}\ .
\end{align}
This is now an equation involving only spacetime points and will be
the starting point for all that follows.

\subsection{Refined duality}

At the moment 
both sides of the equation contain the coupling constant. 
Expanding out the r.h.s. of (\ref{eq:10b}) clearly gives
\begin{align}
  \label{eq:29}
  F_5^{(\ell)}=\sum_{m=0}^{\ell} M_5^{(m)}M_5^{(\ell-m)}\ .
\end{align}
But we can also say something more about the l.h.s. To do this we need
to think a little more graphically than we have so far. In the
previous subsection we reviewed $f$-graphs.
Now to define $F_5^{(\ell)}$ we have done two things, firstly we have
multiplied 
by the external factor $1/f^{(1)} = \prod_{1\leq i<j \leq 5}x_{ij}^2$
and secondly we have taken 
the light-like limit (see~(\ref{eq:28})). Multiplying by  $\prod_{1\leq i<j \leq
  5}x_{ij}^2$ corresponds to deleting all edges between points 1 to 5
(or adding numerator lines if no line exists). Taking the light-like
limit means that any choice of external points 1,2,3,4,5 (recall that
in the $f$-graph we sum over all choices) which are not connected
cyclically via edges $[1,2],[2,3],\dots [5,1]$ will be
surpressed. (Recall an edge $[i,j]$ represents $1/x_{ij}^2$.) So we
only consider as external points, vertices connected in a five-cycle.

Now any cycle on a planar graph immediately splits the graph into
two pieces. E.g. we can embed the graph on a sphere without crossing
(since it is planar) and put the 5-cycle on the
equator thus splitting the graph into a northern and a southern
hemisphere. Alternatively, given an embedding of the graph on the
plane, a 5-cycle splits the graph into an ``inside'' and an
``outside'' graph.

\begin{figure}[!h]
  \begin{tabular}{m{2cm}cm{4cm}cm{4cm}c}
    \includegraphics{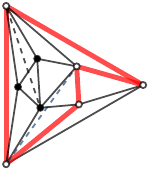}&{\Large $=$}
    & \includegraphics{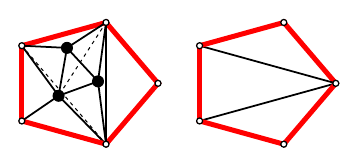}&{\Large$\rightarrow$}
    & \includegraphics{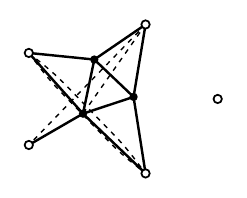}& {\large $\in F_{5;0}^{(3)}$}\\
    \includegraphics{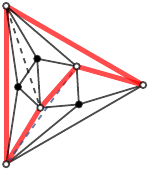}&{\Large$=$}
    & \includegraphics{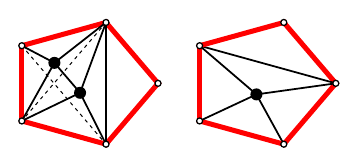} &
    {\Large$\rightarrow$}& \includegraphics{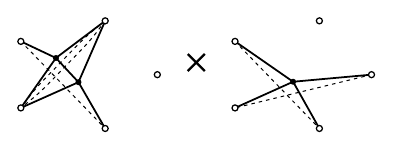} &{\large $\in
      F_{5;1}^{(3)}$}\\$f$-graph with 5-cycle
&& ``Inside''\ \ \ \ \ \ ``Outside''&$ \times 1/f^{(1)}$
\end{tabular}
\caption{Figure illustrating graphically the classification of
      terms in $F_5^{(\ell)}$ into classes $F_{5;m}^{(\ell)}$. We start with a
      single  $f$
      graph (here contributing to $f^{(4)}$,  see~(\ref{eq:6})). The correlator consists
      of summing over all possible 
      labellings of this graph.  Only terms
      where the  external points $1,2,3,4,5$ are consecutively
      connected survive the light-like limit. Such a 5-cycle splits the
      $f$-graph into two pieces, an ``inside'' and an ``outside'' both
      of which are ``disc planar'' i.e. have the right planarity
      properties for amplitude graphs. The minimum of the number of
      vertices inside or outside the 5-cycle gives the value of
      $m$. Here we illustrate with two different 5-cycles (in thick
      red) on the same $f$-graph. The first has $m=0$ and the second
      $m=1$. On the right we give the corresponding amplitude graphs
      ``inside'' and ``outside''.}
  \label{fig:cycles}
\end{figure}

We can now classify terms in $F_5^{(\ell)}$ according to the number $m$ of
points inside (or outside, whichever is smaller)  the corresponding 5-cycle, as
\begin{align}
  \label{eq:30}
  F_5^{(\ell)}= \sum_{m=0}^{\lfloor \ell/2 \rfloor }  F_{5;m}^{(\ell)}\ .
\end{align}
The classification of terms in $F_5^{(\ell)}$ according to their graph
structure is illustrated in Figure~\ref{fig:cycles} 

A simple way of determining the value of $m$ for any given term in
$F_5^{(\ell)}$ is to consider the reduced graph obtained by only
considering edges between internal vertices (i.e. delete all external
vertices). These will in general split into two disconnected groups of size
$m$ and $\ell-m$. 

In any case we see that $F_5^{(\ell)}$ naturally splits into the product of two  graphs just as the duality with the amplitude suggests ($M_5 \bar
M_5$). Note that this split into products
occurs only at the level of the denominator. We can and will see
numerator terms linking the two product graphs. These will be
considered later, but we mention here that such terms are directly
related to parity odd terms in the amplitude.

In summary, then we expect a more refined duality relating specific terms of
$F_5^{(\ell)}$ to specific products of amplitudes as\footnote{Note
  that a completely analogous ``refined'' duality can be given at
  four-points, refining~(\ref{eq:28b}). Namely we define
  $F_{4;m}^{(\ell)}$  as the contribution
  to $F_4^{(\ell)}$ arising from {\em four}-cycles with $m$ points inside
  and $\ell-m$ points outside. Then the refined four-point duality
  reads $ F_{4;m}^{(\ell)} = M_4^{(m)} M_4^{(\ell-m)} $.\label{fn:1}  }
\begin{align}
  \label{eq:31}
  F_{5;m}^{(\ell)}& = M_5^{(m)} \bar M_5^{(\ell-m)} +M_5^{(\ell-m)} \bar
  M_5^{(m)} \qquad m=0 \dots \lfloor (\ell-1)/2 \rfloor\\
  F_{5;m}^{(\ell/2)}& = M_5^{(\ell/2)} \bar M_5^{(\ell/2)} \qquad \ell \in 2 {\mathbb{Z}}\notag
  \ .
\end{align}

For this refined version of the duality to be true as stated we must be certain
there can be no interaction between different terms (i.e. different
values of $m$). The left hand side is clearly well-defined. The inside
and outside of the 5-cycle on a planar f-graph is well-defined. On the
right-hand side we need to ask if all terms in $M_5^{(\ell-m)} \bar
  M_5^{(m)}$ are uniquely identified by their topology as being
  $(\ell-m)$-loops times $m$-loop object. Stated differently, if 
  a pentagon is drawn from points $1,2,3,4,5$ around $M_5^{(m)}$ say,
  can we also draw some or all of $M_5^{(\ell-m)}$ inside the pentagon
  without crossing. One can convince oneself that
  this is indeed not possible: $M_5^{(m)}$ contains at least four
  external vertices, any internal vertex of
  $M_5^{(\ell-m)}$ is connected to at least four external vertices and
  it is impossible to draw two such graphs inside the pentagon without
  crossing. 

\subsection{Four-point graphs appear symmetrically. }

There is a simple all loop consequence of this duality which we
mention here, namely that for 5-point
amplitude graphs depending on only 4 external points (i.e. with one massive
external momentum), the massive point must always appear
symmetrically in all four places (where allowed).

Four-point amplitude graphs only arise in the parity even part of the
amplitude. (The general form of the parity odd part will be discussed
in later sections. Parity odd graphs always depend on all five points.) The
parity even part of the amplitude is given by the  $m=0$ sector of
$F_m^{(\ell)}$ from~(\ref{eq:31}). The $F_m^{(\ell)}$ sector has an
``inside'' and an ``outside'' as discussed in the previous section,
and for $m=0$ the outside (say) has no vertices in it. The outside and
inside must both be planar, but the inside contains a vertex which is
not connected to any other point on the inside (apart from the two
consecutive external points, around the pentagon)  since it supposed to be
a four-point graph. Since the $f$-graph
has degree 4 or more at each point, this means there must be at least
two lines attached to this point on the outside pentagon. The outside
pentagon is then unique given planarity. In other words the ``inside'' and
``outside'' pentagons have the following form which combines into the
$f$-graph on the right. In this picture, the blue edges and vertex
represent the four-point amplitude graph in question (with conformal
weight 1 at all four  points)
\begin{center}
  \includegraphics{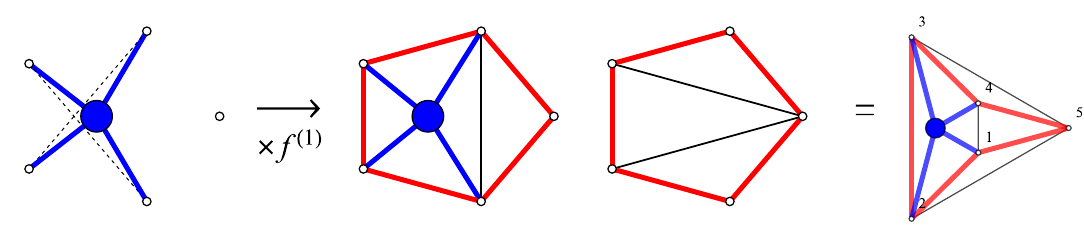}
\end{center}

\vspace{-2.5cm}

\begin{align}
  \label{eq:8}
  \end{align}

\vspace{1cm}
See Figure~\ref{fig:cycles} top row for an explicit example of this.

However, now  we see the $f$-graph this four-point amplitude graph
arises from, we can also see that there are a number of choices of  5 cycles
all giving rise to the same amplitude graph but with the massive leg in
different places:
\begin{center}
  \includegraphics{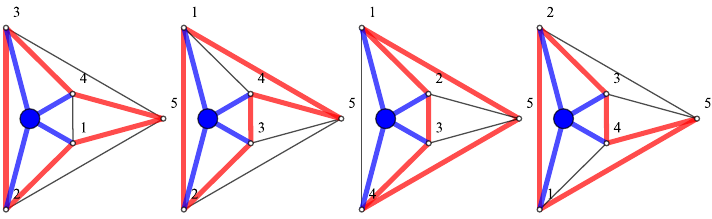}
\end{center}

\vspace{-2.5cm}

\begin{align}
  \label{eq:8a}
  \end{align}

\vspace{1cm}
The massive leg ($x_{14}^2$ in this case)  shifts its position around the
amplitude. We see that any four-point graph will appear symmetrically
with respect to the position of its massive leg in the five-point
amplitude. There is one slightly subtle apparent exception to this
rule. That is the case where the original four-point amplitude has a
numerator term $x_{14}^2$. In this case the numerator means there is
an edge missing in the corresponding $f$-graph and since only one of
the four 5-cycles does not pass through this missing edge, there is only one possible 5-cycle this time as illustrated: 
\begin{center}
  \includegraphics{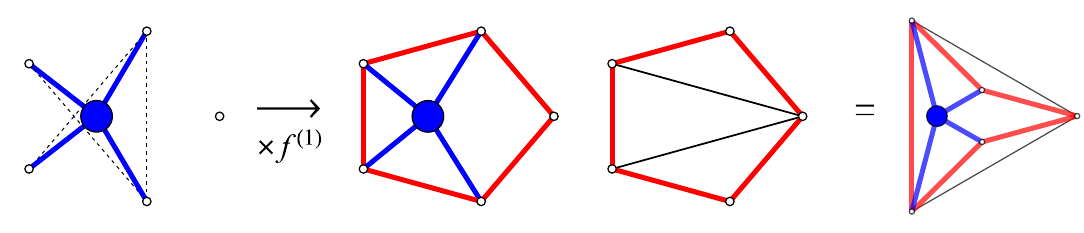}
\end{center}

\vspace{-2.5cm}

\begin{align}
  \label{eq:8}
  \end{align}

\vspace{1cm}
However this is still consistent, since there is also only one allowed
position for the massive leg: all other possibilities will be
suppressed in the light-like limit by this numerator.

In summary, then we find that for any four point topology, the massive
leg appears completely symmetrically. 
For this reason when giving our results we prefer to only display one
representative of this class. We also of course have 5-point
cyclic as well as dihedral symmetry and we only wish to display one
term for all terms related by this symmetry. 

We therefore define an  operator
which we call ``cyc'', which does precisely this, namely cyc[``term'']
denotes the sum over all terms related via cyclic or  dihedral
symmetry, or swapping of the position of the massive  leg in the
four-point case.

We will leave the precise definition of this operation to the
appendix. But suffice it to say here that the  argument of the operation cyc$[]$ always appears with weight 1
when expanding the result into inequivalent terms, i.e.
\begin{align}
  \label{eq:20}
  \text{cyc} [f(x_1,x_2,x_3,x_4)] = f(x_1,x_2,x_3,x_4) + \dots 
\end{align}
where the dots denote different terms.

\section{The one loop five-point amplitude from the correlator}
\label{sec:one-loop-five}

Expanding out~(\ref{eq:10}) to first order in the coupling (equivalent
to 
considering~(\ref{eq:31}) where $m$ can only take the value 0) gives
  \begin{align}
 F_5^{(1)}  \ =\  M_5^{(1)} +\overline{M}_5^{(1)}\ .
\end{align}

The left hand side of this is simply
\begin{align}\label{eq:17}
F_5^{(1)}={\text{cyc}} \left[  \frac{x^2_{13} x^2_{24}}{x^2_{16} x^2_{26} x^2_{36} x^2_{46}} \right] 
\end{align}
which we recognise as the sum over 1 mass boxes. This is indeed twice the
parity even part of the five-point one loop amplitude. 

Having found the parity even part of the one loop amplitude from the correlator,
we now ask if we can obtain the parity odd part? To do so
let us go to next order.

Our refined duality equation~(\ref{eq:31}) with $m=1, \, \ell=2$ gives
  \begin{align}
 F^{(2)}_{5;1}  \ =\  M_5^{(1)}\overline{M}_5^{(1)}\ .\label{eq:7}
\end{align}

So let's check this. The contributions to
$ F_5^{(2)}$ which correspond to product graphs $m=1$  are given
by:
\begin{align}
 F^{(2)}_{5;1} =&\left(\frac{x^4_{13} x^2_{24} x^2_{25}}{x^2_{16} x^2_{17} x^2_{26} x^2_{27} x^2_{36}
    x^2_{37} x^2_{47} x^2_{56}}   \ +\ \notag
  {\text{cyclic in 1,2,3,4,5} } \ + \ x_6 \leftrightarrow x_7\right)\\
&+\frac{x^2_{13} x^2_{14} x^2_{24} x^2_{25} x^2_{35} x^2_{67}}{x^2_{16} x^2_{17} x^2_{26} x^2_{27} x^2_{36} x^2_{37} x^2_{46}
   x^2_{47} x^2_{56} x^2_{57}}\notag\\
=&\,\text{cyc} \left[ \frac{x^4_{13} x^2_{24} x^2_{25}}{x^2_{16} x^2_{17} x^2_{26} x^2_{27} x^2_{36}
    x^2_{37} x^2_{47} x^2_{56}} + \frac{x^2_{13} x^2_{14} x^2_{24} x^2_{25} x^2_{35} x^2_{67}}{x^2_{16} x^2_{17} x^2_{26} x^2_{27} x^2_{36} x^2_{37} x^2_{46}
   x^2_{47} x^2_{56} x^2_{57}} \right]
\end{align}
Equating this to $M_5 \bar M_5$, together with~(\ref{eq:17}) gives us two
equations for two unknowns,  $M_5$ and  $\bar M_5$ and we can thus
solve for them. The equations are quadratic and so the
solution 
involves a
square root whose sign we will not be able to determine without more information.

The solution is simply 
\begin{align}
  M_5^{(1)}
&={1 \over 2} \Big( F_5^{(1)} \pm
  \sqrt{(F_5^{(1)})^2 -4 F^{(2)}_{5;1}}\Big)
\\\overline{M}_5^{(1)} &= {1 \over 2} \Big( F_5^{(1)} \mp
  \sqrt{(F_5^{(1)})^2 -4 F^{(2)}_{5;1}}\Big)\ .
\end{align}
We have written the full parity even and odd 5-point ampitudes in
terms of purely parity even objects (but involving a square root).

One can now ask if there is a better way of writing the parity odd part of
this without using the square root, and indeed this is the case.

There is a unique parity odd conformally invariant tensor, which is
easiest to see in the six-dimensional formalism reviewed in Appendix B.
In this formalism it is clear that there is a unique parity odd
conformally covariant object. It is a function 
of six
points, $x_1, \dots x_6$, each with  weight 1 which we denote
$\epsilon_{123456}$. It has a natural form in the six-dimensional
formalism, but can be written in various different ways in standard
four-dimensional formalism (see
section~\ref{sec:diff-forms-epsil}). In any case using this object one can show that the term inside the
square root (thought of as an integrand product with integrand points
$x_6$ and $x_7$ which are symmetrised) can be written in the more
suggestive form
\begin{align}
 (F_5^{(1)})^2 -4 F^{(2)}_{5;1} =
 - {\epsilon_{123456} \over
    x^2_{16}x^2_{26}x^2_{36}x^2_{46}x^2_{56} } \times
  {\epsilon_{123457} \over x^2_{17}x^2_{27}x^2_{37}x^2_{47}x^2_{57} }\ .
\end{align}
To see this, use the identity 
\begin{align}
 & \epsilon_{123456} \times   \epsilon_{123457} \notag\\
=&  \text{cyc}\Big[ 2 \, x_{67}^2
    x_{13}^2 x_{24}^2 x_{35}^2 x_{14}^2 x_{25}^2 + x_{13}^4 x_{24}^2 x_{25}^2 x_{46}^2 x_{57}^2 - x_{13}^4x_{24}^4 x_{56}^2 x_{57}^2
   - x_{13}^2 x_{14}^2 x_{24}^2 x_{25}^2 x_{36}^2 x_{57}^2   \Big]\, .  \label{eq:15}
\end{align}

\begin{figure}[!ht]
\label{fig:1}
  \centerline{
 \includegraphics{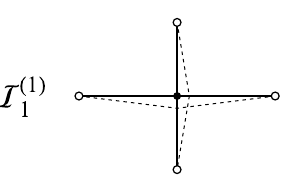} \qquad
 \includegraphics{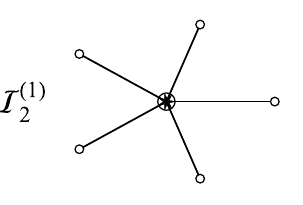}}
  \caption{One loop five-point parity even and odd amplitude graphs. This is
    just a one loop box in dual coordinates,  and a pentagon
    graph. The starred vertex $v$  indicates a factor $i
    \epsilon_{12345v}$. }
\end{figure}

We then obtain our final result for the five-point amplitude to be 
\begin{align}\label{eq:24}
  M_5^{(1)} =\frac12 \left( \cI_1^{(1)} +\cI_2^{(1)} \right)\ .
\end{align}
The terms in this amplitude are displayed graphically in
figure~\ref{fig:1}.

\begin{equation}
 \input{1loopf.tex}\label{eq:23}
 \end{equation}

In section 7 we show that this form of the five-point amplitude integrand is consistent with
both the local expression in terms of twistors \cite{nima2}, and with the all orders in $\epsilon$
version containing a parity odd pentagon at order $\epsilon$, \cite{hep-th/9611127}.

\section{Two loops}
\label{sec:two-loops}

We now proceed to investigate $M_5^{(2)}$. 
The refined duality equation~(\ref{eq:31}) gives two equations
involving $M_5^{(2)}$ and lower loop amplitudes, namely for $\ell=2,m=0$
and for $\ell=3,m=1$
\begin{align}
{F^{(2)}_{5;0} }  \ &=\ M_5^{(2)}+ \overline{M}_5^{(2)}\\
{F^{(3)}_{5;1} }  \ &=\  M_5^{(2)} \overline{M}_5^{(1)}+M_5^{(1)} \overline{M}_5^{(2)}\ .
\end{align}
Therefore as before, since we have two equations for two unknowns, $M_5^{(2)}$
and $\overline{M}_5^{(2)}$, we can solve for these.

To do this first rewrite the equations as:
\begin{align}
M_5^{(2)}+ \overline{M}_5^{(2)}  \ &=\ {F^{(2)}_{5;0} }\\
  \left(M_5^{(2)}-\overline{M}_5^{(2)}\right)\left(M_5^{(1)}-\overline{M}_5^{(1)}\right)
&= F^{(2)}_{5;0} F^{(1)} -2
   F^{(3)}_{5;1}\ ,\label{eq:11}
\end{align}
thus giving an equation for the parity odd part of the two loop
amplitude in term of correlator quantities $F$'s and the one loop
parity odd amplitude.

Once more we can simplify the parity odd part of the amplitude at two
loops. To do this, we write an ansatz for the form of
$M_5^{(2)}-\overline{M}_5^{(2)}$. Since it is parity odd it must
contain one factor of the six-dimensional $\epsilon$ tensor. By
examination we find the parity odd part of the two loop amplitude is
\begin{align}
  {M_5^{(2)}-\overline{M}_5^{(2)}} = {1\over 2!}\text{cyc}\left({\pm i \epsilon_{123456}
    x^2_{35}\over
    x^2_{16}x^2_{26}x^2_{36}x^2_{56}x^2_{37}x^2_{47}x^2_{57}x^2_{67}} \right)
\end{align}
which is a pentabox with an epsilon in the numerator. Note that the
$\pm$ here is the same as the 1 loop one, so once that sign is fixed
so will this two loop one.

The full two-loop amplitude is then
\begin{align}
  \label{eq:22}
  M^{(2)}_5 = \frac1{2 \times 2!} \left( \cI_1^{(2)}+ \cI_2^{(2)}+ \cI_3^{(2)}\right)
\end{align}
where
\begin{equation}
 \input{2loopf.tex}
\end{equation}
with corresponding graphs
\begin{figure}[!ht]
  \centerline{
  \includegraphics{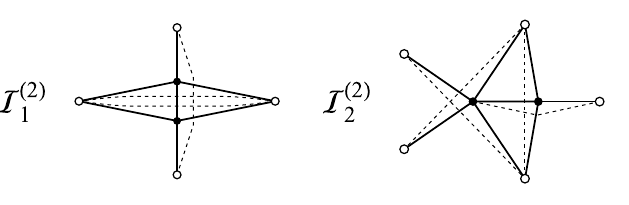}
\quad   \includegraphics{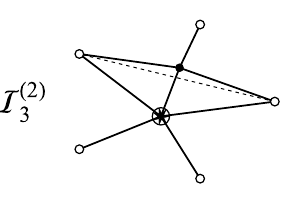}}
  \caption{Two loop five-point parity even ($\cI_1^{(2)}$ and
    $\cI_2^{(2)}$) and parity odd ($\cI_3^{(2)}$) amplitude graphs.  The starred vertex $v$  indicates a factor $i
    \, \epsilon_{12345v}$. }
\end{figure}

\section{Higher loops}

This process can clearly be extended to higher orders. At $\ell$-loops
we use the refined duality~(\ref{eq:31}) with $\ell,m=0$ and
$\ell+1,m=1$ giving
\begin{align}
{F^{(\ell)}_{5;0} }  \ &=\ M_5^{(\ell)}+ \overline{M}_5^{(\ell)} \label{eq:4}\\
{F^{(\ell+1)}_{5;1} }  \ &=\ M_5^{(\ell)} \overline{M}_5^{(1)}+M_5^{(1)}
\overline{M}_5^{(\ell)} \ .\label{eq:9}
\end{align}

From~(\ref{eq:4}) we can immediately read off the
parity even part $ M_5^{(\ell)}+ \overline{M}_5^{(\ell)}$. Then similarly to~(\ref{eq:11}) we can write
\begin{align}
  \left(M_5^{(\ell)}-\overline{M}_5^{(\ell)}\right)\left(M_5^{(1)}-\overline{M}_5^{(1)}\right)
&= F^{(\ell)}_{5;0} F^{(1)} -2
   F^{(\ell+1)}_{5;1}
\ ,\label{eq:11b}
\end{align}
giving the parity odd part of the $\ell$ loop graph in terms of
correlator quantities ($F$'s) and the one-loop amplitude. 
So knowing the right-hand side of this equation we can compute the
parity odd combination $M_5^{\ell}-\bar M_5^{(\ell)}$.

Now as at two loops we wish to rewrite this in a simpler form, i.e. in
terms of $\epsilon_{123456}$. In principle we could include epsilon
objects with two or more internal variables so for example $\epsilon_{123467}$.
However we have always  found solutions in which only a
single internal variable appears in the $\epsilon$. We therefore make the following assumption:
\vspace{.5cm}

{\bf Assumption:} The parity odd part of the
five-point amplitude at any loop can always be written in the form $\int d^4 x_6\dots d^4x_{5+\ell} \,
\epsilon_{123456} \, f(x_i)$  where $f(x_i)$ is an integrand composed of $x_{ij}^2$ 
depending on all external and internal variables. There never is an epsilon tensor involving two or more internal points.

\vspace{.5cm}


With the help of this it is remarkably straightforward to
compute the
parity odd part of the amplitude at $\ell$ loops from the correlator. 
In the combination
$
\left(M_5^{(\ell)}-\overline{M}_5^{(\ell)}\right)\left(M_5^{(1)}-\overline{M}_5^{(1)}\right)$
on the l.h.s. of~(\ref{eq:11b}) we have to consider the product of two epsilon tensors, one from
$\ell$ loops using the above conjecture and one from one loop. This
product 
contains a single  term involving an inverse propagator between two internal
vertices (see (\ref{eq:15}))
\begin{align}
  \label{eq:12}
  \epsilon_{123456} \, \epsilon_{123457} \, = \, 2 \, {\bf x_{67}^2} \, x_{13}^2x_{35}^2x_{25}^2x_{24}^2x_{14}^2
  + \dots\ .
\end{align}
Thus this will produce a product graph, a pentagon around $x_6$ glued
to a higher loop graph involving $x_7$ together with a numerator
$x_{67}^2$ between them. Such a product graph with numerator can be produced from
the correlator $F^{(\ell+1)}_{5;1}$ but can not be
cancelled by any 
terms on the right hand side of~(\ref{eq:11b}). Thus each graph of this
type in $F^{(\ell+1)}_{5;1}$
uniquely singles out a corresponding $\epsilon$-term in $
M_5^{(\ell)}-\overline{M}_5^{(\ell)}$. 

This can again be interpreted in terms of correlator
$f$-graphs:  5-cycles in the $f$-graph  split the
graph into two halves. We look for 5-cycles which have the
1 loop pentagon graph on one side. The other side then gives us the
parity odd graph in question. Its coefficient is inherited from
the $f$-graph. The procedure is illustrated in Figure~4.
\begin{figure}[!h]\label{fig:3}
  \centering
  \includegraphics{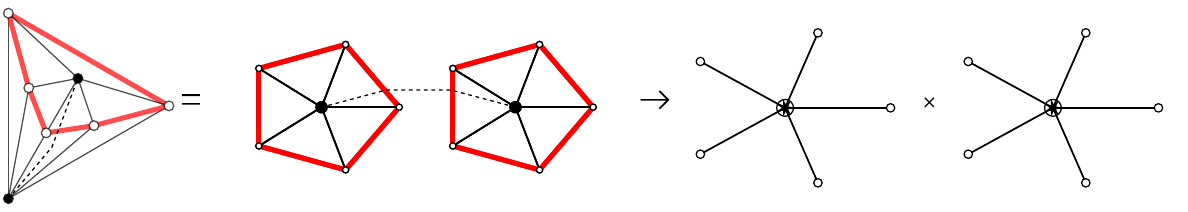}  \includegraphics{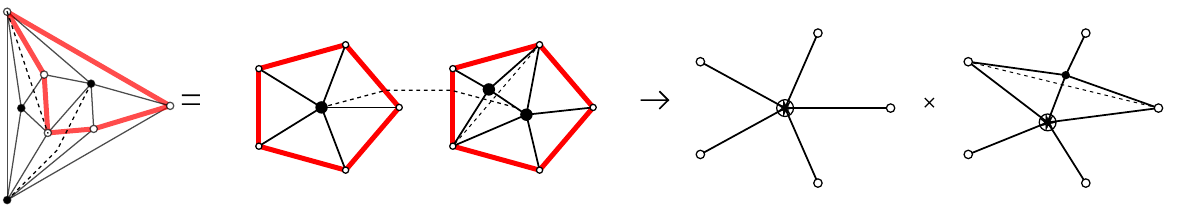}  
  \caption{Figure illustrating the procedure for obtaining the parity
    odd part of the five-point amplitude from the correlator
    $f$-graphs. The 5-cycle (shown in thick red) splits the graph into two parts. The inside of
    the 5-cycle corresponds to the 1 loop parity odd pentagon, whereas
    the outside corresponds to the higher loop parity odd graph. The
    starred vertex is the vertex attached to the 1 loop internal
    vertex via an internal line. In the first line we start with a
    5-cycle in 
    $f^{(3)}$ contributing to $F_1^{(2)}$, the ``outside'' of which determines the parity odd
    graph for $M_5^{(1)}$. In the second line we start with 
    a 5-cycle in one of the
    three $f$-graphs contributing to $f^{(4)}$
    contributing to $F_5^{(2)}$ thus giving a contribution to
    $F^{(3)}_{5;1}$. The ``outside'' of the 5-cycle then determines the
    parity odd 
    graph for $M_5^{(2)}$.}
\end{figure}

That this simple rule then correctly reproduces the entire right-hand side of~(\ref{eq:11b})
appears somewhat miraculous and relies on many cancellations between graphs. { We will attempt to give some
  motivation of why/how this works in the conclusions. Notice that
  this consistency determines many of the correlator coefficients not determined
  from the four-point duality (determined by the rung rule which
  arises from consistency of the four-point amplitude/correlator duality). The first
  coefficient not determined by five-point consistency appears in $f^{(6)}$.}

Note there are of course further consistency requirements on this picture,
starting at four loops, since we require the $m=2$ part of $F_5^{(4)}$
to be given by the product of two loop amplitudes (which were
determined by $F^{(2)}_{5;0}$ and $F^{(3)}_{5;1}$  i.e. $F^{(4)}_{5;2}=M_5^{(2)}\bar
M_5^{(2)}$.

\begin{figure}[H]
  \centerline{
  \includegraphics{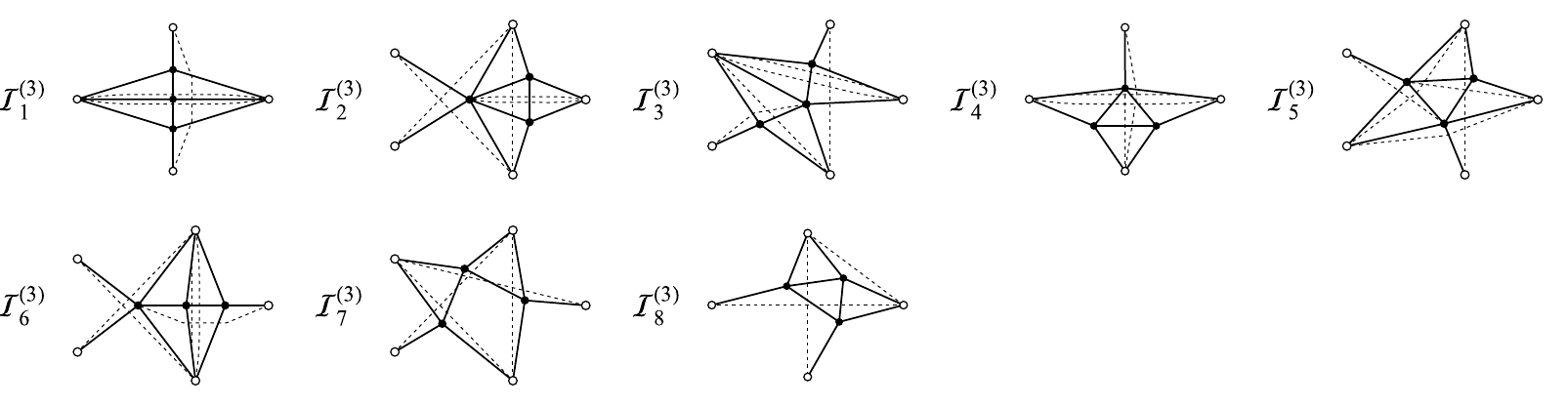}}
  \caption{Three loop five-point parity even amplitude graphs. White
    vertices indicate external points, $x_1,x_2,x_3,x_4,x_5$ which are
    cyclically ordered in either orientation and black
    nodes integration variables $x_6,x_7,x_8$.  A black
  line between vertices $i$ and $j$ indicates a propagator term
  $1/x_{ij}^2$. A dotted line indicates an inverse propagator
  $x_{ij}^2$. The integrand expression $\cI_i^{(3)}$ is the expression
  thus obtained by summing over all {\em different}
  such labellings. 
}
\label{fig:3loopeven}
\end{figure}

\begin{figure}[H]
  \centerline{
  \includegraphics{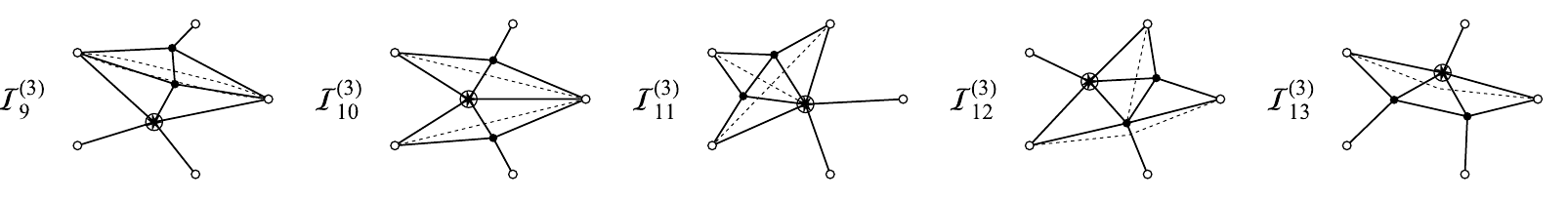}}
  \caption{Three loop five-point parity odd amplitude graphs. A
    starred vertex $v$ indicates a factor $i \epsilon_{12345v}$.}
\label{fig:3loopodd}
\end{figure}

Using this method we have obtained the full the three-loop five-point amplitude (parity even and parity odd part) and checked that it indeed satisfies the consistency condition~(\ref{eq:11b}):
\begin{align}
  \label{eq:13}
  M_5^{(3)}= {1 \over 2} {1\over 3!} \int d^4x_6 d^4x_7 d^4x_8 \left(
    \sum_{i=1}^{13} c_i \cI_i^{(3)} \right)\, ,
\end{align}
where 
\begin{align}
  \label{eq:14}
  c_1=\dots =c_6= c_9=\dots c_{12}=1\, , \qquad c_7=c_8=c_{13}=-1 .
\end{align}
and 
\begin{equation}
  \label{eq:16}
\input{3loopf.tex}
\end{equation}
also illustrated graphically in figures~\ref{fig:3loopeven} and~\ref{fig:3loopodd}

\section{Four- and five-loops amplitude}

Similarly, using the method outlined in the previous section we have obtained the full (parity even and parity odd part) four-loop five-point amplitude and checked that it
satisfies the consistency condition~(\ref{eq:11b}).
For the four-loop result:

\begin{align}
  \label{eq:13b}
  M_5^{(4)}={1 \over 2} {1\over 4!} \int d^4x_6 d^4x_7 d^4x_8 d^4x_9 \left(
    \sum_{i=1}^{71} c_i \cI_i^{(4)} \right)\, ,
\end{align}
where 
\begin{align}
  \label{eq:14b}
  c_1=\dots =c_{28}= c_{45}=\dots = c_{62}=1 \, , \\ c_{29}=\dots
  =c_{44}=c_{63}=\dots=c_{71}=-1 \ \notag
\end{align}

and 
{\tiny \begin{align*}
 \input{4loopf.tex}
\end{align*}
}%

\begin{figure}[H]
  \centerline{
  \includegraphics{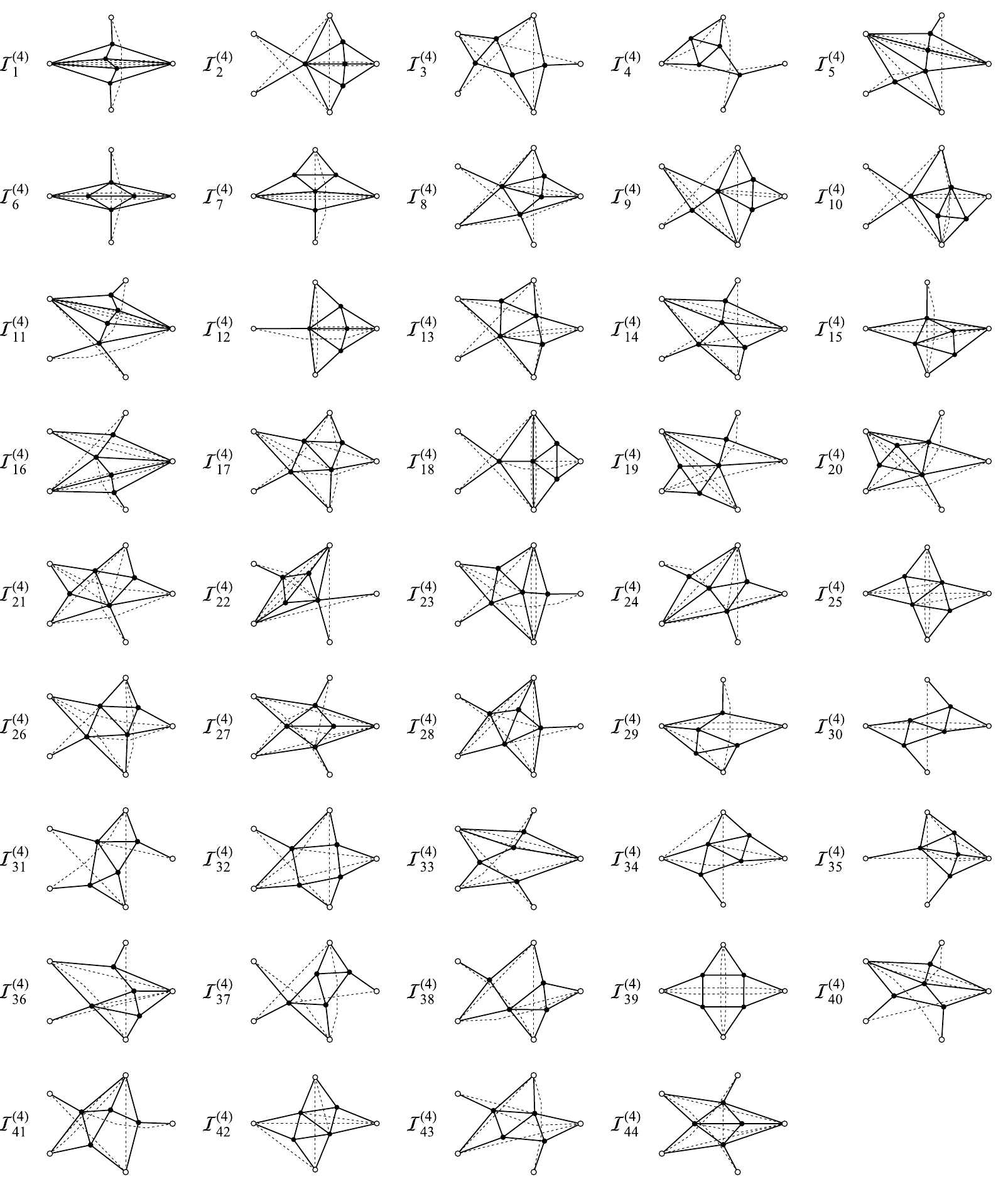}}
  \caption{Four loop five-point parity even amplitude graphs.}\label{fig:4le}
\end{figure}

\begin{figure}[H]
  \centerline{
  \includegraphics{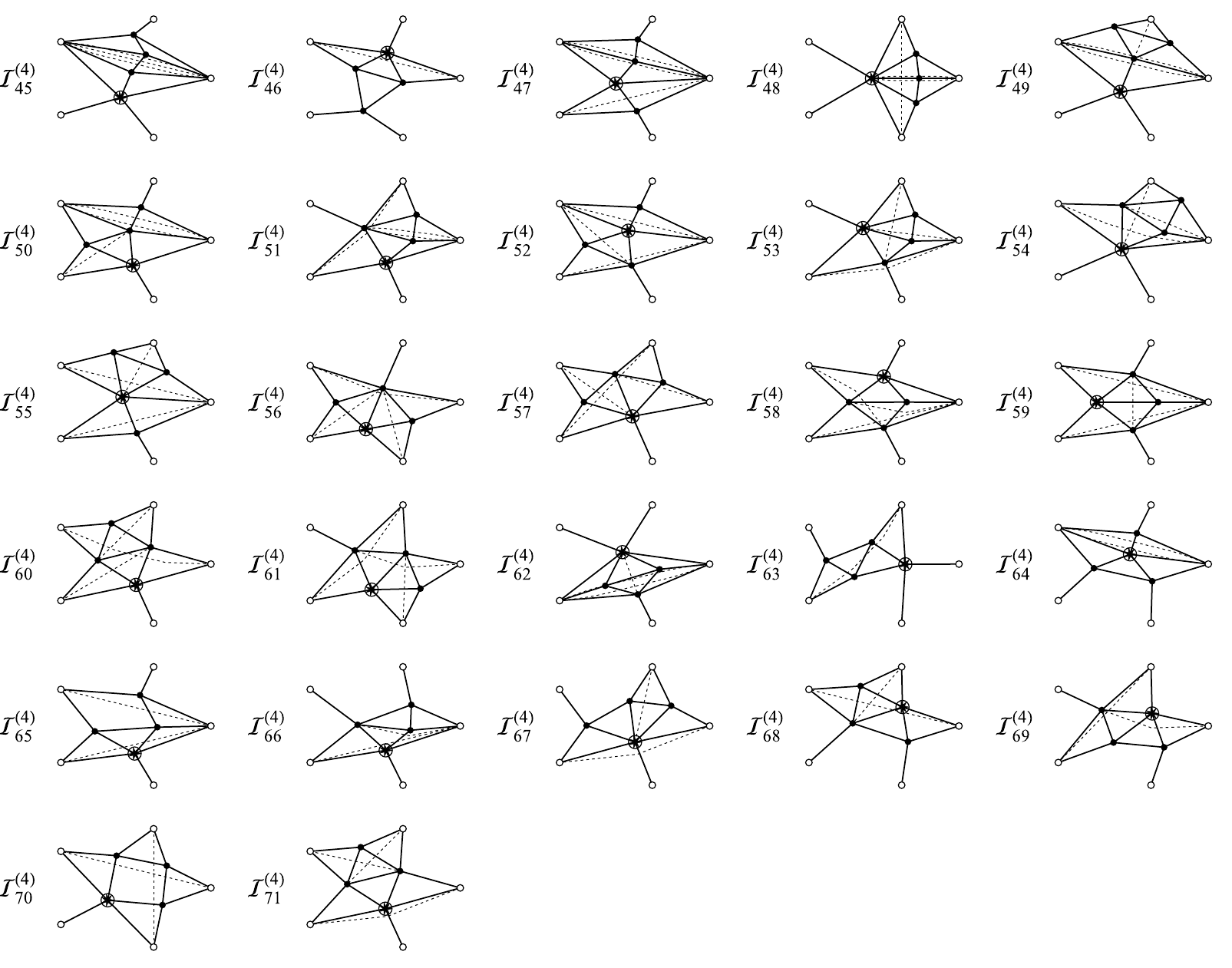}}
  \caption{Four loop five-point parity odd amplitude graphs. A
    starred vertex $v$ indicates a factor $i \epsilon_{12345v}$.}\label{fig:4lo}
\end{figure}

The corresponding four-loop amplitude graphs are given in Figures~\ref{fig:4le} and~\ref{fig:4lo}.

We have also been able to obtain the full five-loop parity even and
odd amplitude. In order to do this we needed $f^{(7)}$ which was
obtained in~\cite{charles} from the four-point seven-loop
amplitude~\cite{1112.6432}. The seven-loop $f$-graphs and their coefficients are contained in the
two separate files \texttt{7LoopTopologies.txt} and \texttt{7LoopCoefficients.txt} attached to the arXiv version of this paper. The result for the five-loop five-point amplitude consists of 318 different parity even topologies and 203 parity odd graphs which we give in the file \texttt{5pointamplitude.txt}, which also contains the six-loop parity even integrand. As a piece of complementary information \texttt{5pointamplitudenumberofterms.txt} contains the number of independent terms obtained from every graph in \texttt{5pointamplitude.txt} by the cyc$[]$ operation.
In order to obtain the parity odd part of the six-loop amplitude we would need
$f^{(8)}$ which could be obtained for example directly from the four-point eight-loop
amplitude if it became available.

\section{Relation to other ways of writing 5-point integrands}

\subsection{Momentum space integrands}
\label{sec:scalar-integrands}

Forms for 1- and 2-loop 5-point amplitudes are available in momentum space in the literature both in the planar~\cite{hep-th/9611127,hep-th/0604074} and more recently the full non-planar theory~\cite{1106.4711}. Whilst it is straightforward to write our dual momentum space integrands in terms of momentum space (simply using the replacement $p_i=x_{i\,i+1}$) a direct comparison requires some manipulation. In particular the preferred bases  have integration variables appearing only as scalar products, rather than in parity odd epsilon tensors as we have here. The parity odd epsilon tensors then only depend on external momenta. 

We can easily rewrite our integrands in such a form using a single formula for rewriting $\epsilon_{123456}$ in terms of $x_{i6}^2$  derived in
appendix~\ref{sec:diff-forms-epsil} namely
\begin{align}
  \epsilon_{123456}
&=  \frac{s_2 s_3}{8 \,\epsilon(p_1,p_2,p_3,p_4)}
 Tr(\slashed{p}_4 \slashed{p}_5 \slashed{p}_1 \slashed{p}_2)
 \,x_{16}^2 \quad + \ \text{cyclic 1,2,3,4,5}\notag\\
 &\qquad  +    \frac{s_1s_2s_3 s_4
   s_5}{16\,\epsilon(p_1,p_2,p_3,p_4)}\ ,\label{eq:41}
\end{align}
where $p_i=x_{i\,i+1}$ are the amplitude momenta and $
s_i=x_{i\,i+2}^2 = (p_i+p_{i+1})^2$ the usual two-particle invariants. Note that this form breaks manifest dual conformal symmetry, and the
coefficients of the integrands are more ugly, but it enables fairly direct comparison with results in the literature. 

The canonical example is one loop. From~(\ref{eq:24}) the 1 loop
parity odd term is
$$ \frac{i \epsilon _{123456}}{x_{16}^2 x_{26}^2
    x_{36}^2 x_{46}^2 x_{56}^2}$$
which, using~(\ref{eq:41}) becomes
\begin{align}
  \label{eq:42}
  & \frac{i \,s_2 s_3} {8 \,\epsilon(p_1,p_2,p_3,p_4)}
 Tr(\slashed{p}_4 \slashed{p}_5 \slashed{p}_1 \slashed{p}_2)
 \,\times\,\text{Box}(p_2,p_3,p_4,p_5+p_1)
 \quad + \ \text{cyclic 1,2,3,4,5}\notag\\
 &\qquad  +    \frac{i\,s_1s_2s_3 s_4
   s_5}{16\,\epsilon(p_1,p_2,p_3,p_4)}\,\times\,\text{Pent}(p_1,p_2,p_3,p_4,p_5)
\end{align}
where we see the (one mass)   scalar box integrands and a scalar pentagon
integrand
\begin{align}
  & \text{Box}(p_2,p_3,p_4,p_5+p_1)= \int d^4x_6 \frac1{x_{26}^2x_{36}^2x_{46}^2x_{56}^2} \, , \\
  & \text{Pent}(p_1,p_2,p_3,p_4,p_5) = \int d^4x_6 \frac1{x_{16}^2 x_{26}^2x_{36}^2x_{46}^2x_{56}^2}
\end{align}
with non-trivial coefficients. These can then be
compared directly with for example with the form for non-planar
amplitudes (in the 
planar limit) found in~\cite{1106.4711} where the same basis of scalar
integrals is used and we find
perfect matching. 

At two loops, using the same formula we reproduce the form of the amplitude given in~\cite{hep-th/0604074}.

\subsection{Twistor space integrands}
\label{sec:twistor-integrands}

General expressions for MHV amplitude integrands up to three loops
(parity even and odd) have also been given in 
momentum twistor space~\cite{nima2}. We will not review
momentum twistors here. The only information we will need is the relation
to the six-dimensional formalism reviewed in
appendix~\ref{sec:4d-mink-coord}.
When the  $X_{i}$ are consecutively lightlike separated (i.e. $X_{i}
\cdot X_{i+1}=0$) then we let ${X}_{i}^{AB} = Z_{i-1}^{\left[A \right.}
Z_{i}^{\left. B \right]}$. Now let's consider various integrands as
they are expressed in~\cite{nima2}. All one-loop integrands are
written in terms of a dual-conformal pentagon integral:
\begin{equation}
I_{i,j} = \frac{ \langle AB (i-1,i,i+1) \cap (j-1,j,j+1) \rangle\langle X,i,j \rangle}{\langle AB, i-1,i \rangle \langle AB, i,i+1 \rangle \langle AB, j-1,j \rangle \langle AB,j,j+1 \rangle \langle ABX \rangle}
\end{equation}
This can be rewritten in terms of a
trace over $X$ variables
\begin{equation}
I_{i,j} = \frac{\mathrm{Tr}(X \tilde{X}_{i+1} X_{i} \tilde{X}_{0} X_{j}
  \tilde{X}_{j+1})}{(X_{0} \cdot X_{i})(X_{0} \cdot X_{i+1})(X_{0}
  \cdot X_{j})(X_{0} \cdot X_{j+1})(X_{0} \cdot
  X)} \label{5pointoneref}\ .
\end{equation}
Here the integration over twistors $A$, $B$ has become integration over
the $X$-space variable $X_{0}$. The variable $X$ is a reference
twistor meaning it should drop out of the
sum which gives the one-loop amplitude. Indeed the simplest way to
deal with it is to set it to be one of the external points, e.g.
$X=X_5$.
So the one loop amplitude can be written entirely in terms of a
six-trace. Just as for the more familiar four-traces in QFT, the six-trace
splits as a parity odd term $\epsilon$ and a parity even term
(multiple scalar products)
\begin{align}
  \label{eq:44}
  \text{Tr}(X \tilde{X}_{i+1} X_{i} \tilde{X}_{0} X_{j}
  \tilde{X}_{j+1})) = 4\, i \, \epsilon_{X(i+1)i0j(j+1)} +4\, (X_0.X)(X_{i+1}.X_{j+1})(X_i.X_j) + \dots
\end{align}
where the dots indicate similar terms which can be obtained by taking
all possible combinations of scalar products between the six entries
of the trace with minus signs where appropriate.  We see immediately that the parity odd
part of this twistor integrand will yield exactly the same pentagon $\cI_2^{(5)}$ derived from the correlator. After summing all diagrams, the parity even terms will also give the same sum over one-mass boxes we expect.

Similarly at two loops, the $n$-point momentum twistor integrand in~\cite{nima2} is
\begin{align}
M_5 \, = \, &{{\frac{1}{2} \sum_{i<j<k<l<i} \frac{ \langle AB (i-1,i,i+1) \cap
      (j-1,j,j+1) \rangle \langle i,j,k,l \rangle }{\langle AB,i-1,i
      \rangle \langle A B i,i+1 \rangle \langle AB,j-1,j \rangle
      \langle AB, j,j+1 \rangle \langle ABCD \rangle}}} \notag\\
& \qquad \qquad \qquad \qquad \qquad \times \frac{\langle C D (k-1,k,k+1) \cap (l-1,l,l+1) \rangle}{ \langle CD,k-1,k
      \rangle \langle CD,k,k+1 \rangle \langle CD,l-1,l \rangle
      \langle CD,l,l+1 \rangle}
\label{2loopX}
\end{align}
which we can re-write in $X$-coordinates (the
integration variables are $AB=X_0$,  $CD=X_{\bar 0}$) in the form:
\begin{equation}
\frac{1}{2} \sum_{ijkl} \frac{\left[ X_{i+1} \tilde{X}_{i} X_{0}
    \tilde{X}_{j+1} X_{j} \right]^{AB} \left[ X_{k+1}
    \tilde{X}_{k} X_{\bar 0} \tilde{X}_{l+1} X_{l} \right]^{CD}
  \epsilon_{ABCD}}{(X_{0} \cdot X_{i})(X_{0}
  \cdot X_{i+1})(X_{0} \cdot X_{j})(X_{0} \cdot X_{j+1})(X_{0}
  \cdot X_{\bar 0})(X_{\bar 0} \cdot X_{k})(X_{\bar 0} \cdot
  X_{k+1})(X_{\bar 0} \cdot
  X_{l})(X_{\bar 0} \cdot X_{l+1})}
\end{equation}
 To rewrite further we take advantage of 'boundary cases',
 i.e. at 5-points either $j=i+1$ or $j=i+2$ and $l=k+1$. For example
 when $j=i+1$  we get:
\begin{equation}
 \frac{X_i.X_{i+2} \, \mathrm{Tr}(X_{\bar 0} \tilde{X}_{k} X_{k+1}
   \tilde{X}_{i+1} X_{l+1} \tilde{X}_{l})}{(X_{0} \cdot X_{i})(X_{0}
   \cdot X_{i+1})(X_{0} \cdot X_{i+2})(X_{0} \cdot X_{\bar 0})(X_{\bar 0}
   \cdot X_{k})(X_{\bar 0} \cdot X_{k+1})(X_{\bar 0} \cdot X_{l})(X_{\bar 0} \cdot
   X_{l+1})}
\end{equation}
In this way and using~(\ref{eq:44}) we indeed recover the two loop 5-point amplitude in the
form (\ref{eq:22}).

At three loops, starting from the equations given in~\cite{nima2} we are able to reproduce the same set of graphs which we have produced here. The mapping for parity odd terms is very simple and can be seen directly from drawing graphs dual to those given in~\cite{nima2}, however the parity even terms are significantly more complicated.

\section{Conclusions}

The supersymmetric correlator/amplitude duality in ${\cal N} = 4$ gives a way of relating objects with different numbers of outer points, or in- or outgoing particles, respectively. In the present article we have exploited this feature of the construction to derive the integrand of the colour ordered five-point amplitude up to five (and in the parity even sector six) loops from that of the four-point function of energy-momentum multiplets, which was so far chiefly associated with the MHV four-point amplitude \cite{hidden,constructing}.

In order to take the step from four to five points, one of the integration vertices of the four-point integrand has to be regarded as an outer point. Necessarily we lose one loop order in this way. It turns out that the five-point integrand can only be uniquely fixed by taking into account topological information: amplitude graphs are planar on the disc, while the correlator integrands also contain products of two such graphs. We have used the one-loop $\times$ higher-loop terms to gain more equations on the loop corrections to the five-point amplitude. Stripping off a one-loop amplitude implies losing another loop order, though.

A beautiful picture then emerges where the parity even five-point
$\ell$-loop amplitudes correspond to the outsides of those five-cycles in the
planar correlator $f^{(\ell+1)}$-graphs which have no vertices on
their insides, whereas the
parity odd amplitude graphs correspond to the outsides of those five-cycles in the
planar correlator $f^{(\ell+2)}$-graphs which have a single  vertex on
their inside.

Our main new results are the four- and five-loop integrands for the five-point
MHV (or in this case equivalently the NMHV) amplitude. To this end,
the analysis of \cite{constructing} was extended to the seven-loop
integrand of the four-point correlation function of energy-momentum
multiplets based on the result \cite{1112.6432} for the four-point
MHV amplitude up to seven loops. We have thus made a four-point into a
five-point amplitude.

Indeed, that this picture works out to be consistent is rather remarkable and
non-trivial. The duality with four-point
amplitudes can be shown to be consistent as long as the corresponding
amplitude graphs obey the  rung rule~\cite{hep-ph/9702424} which in the correlator picture
simply corresponds to gluing pyramids onto the $f$-graphs~\cite{constructing}.  Indeed  the mere
existence of the four-point duality then predicts many of the
coefficients of loop level amplitudes (all up to three loops, the first
two out of
the three four-loop $f$-graphs, and the first six out of seven
five-loop $f$-graphs (see~(\ref{eq:6}) etc.) What is the topological reason stopping certain four-point $f$-graphs
being determined from lower loops? Recall the  refined four-point duality (see
 footnote~\ref{fn:1}) $2F_{4;m}^{(\ell)}=M_4^{(m)}M_4^{(\ell-m)}$. Thus
$f$-graphs with four-cycles with  a non-trivial ``inside'' and
``outside'' (i.e. which contribute to $m>0$) are determined entirely in
terms of lower loop amplitudes. Conversely $f$-graphs which  give no
contribution to $F_m^{(\ell)}$  for $m>0$, i.e. which have no such
four-cycle, cannot be determined from lower loop four-point
amplitudes (see the final two graphs in $f^{(4)}$ and $f^{(5)}$ in~(\ref{eq:6})).

For the five-point duality on the other hand
the consistency is much more subtle and we have no clear understanding
(i.e. a generalisation of the pyramid gluing rung rule) for why this
works.
The confusion comes from the many terms which appear when gluing two
$\epsilon_{123456}$ together, many of which have to cancel.  However
we have 
noticed that the structure does indeed determine many of the
non-rung-rule-determined coefficients. Indeed merely the structure and
consistency of the picture determines all coefficients up to
$f^{(5)}$, i.e. the mere existence of the amplitude/correlator duality
at 4- {\em and} 5-points determines the four-point correlator and amplitude
to five loops and the five-point amplitude to four loops (parity even)
and three loops (parity odd). The first  coefficient which is not
determined by these purely structural arguments is that of the
10-point (6 loop)  $f$-graph:
\begin{center}
  \includegraphics[width=2cm]{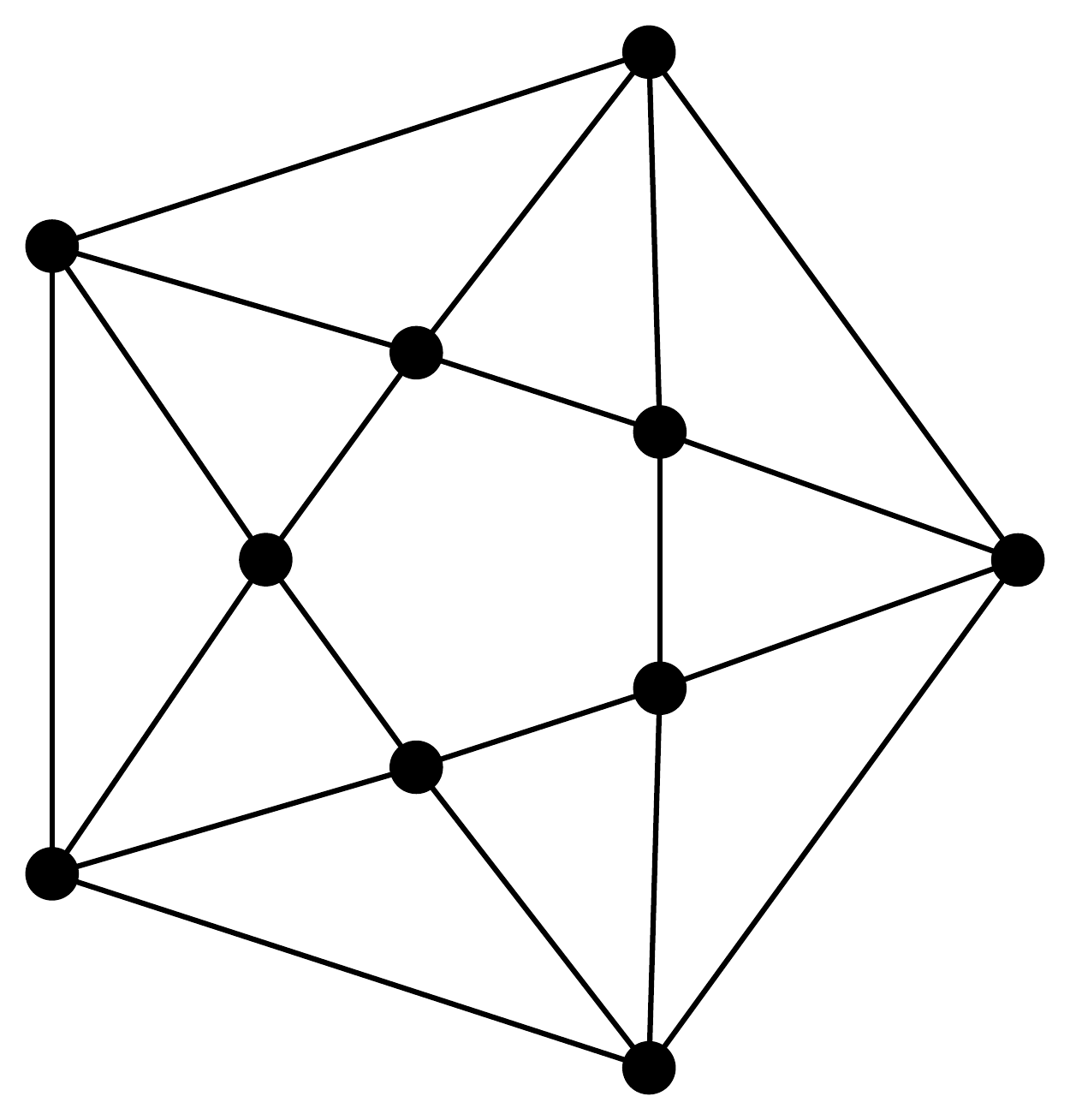}
\end{center}
Clearly any $f$-graph giving no  contribution to $F_{5;m}^{(\ell)}$ for
$m>0$ (i.e. 
whose 5-cycles have either no vertices inside or none outside) will
not be determined by lower loops and it seems likely that the converse
is true also: any $f$-graph contributing to $F_{5;m}^{(5)}$ for
$m>0$ will be determined from lower loops via the refined
duality~(\ref{eq:31}) $F_{5;m}^{(\ell)} = M_5^{(m)} \bar
M_5^{(\ell-m)} +M_5^{(\ell-m)} \bar  M_5^{(m)}$.\footnote{Here it is a
  little  subtle since we only determine the parity odd part of
  $M_5^{(\ell-1)}$ from $F^{(\ell)}$ itself. However the parity even
  part also contributes to this formula, and so unless there is
  complete cancellation between parity even and parity odd, which seems
unlikely, $F_{5;m}^{(\ell)}$ and the corresponding $f$-graph will be determined by the lower loop
amplitude. } 
Indeed we see that all the 5-cycles of the graph above have
either nothing inside or nothing  outside them and this is the first
such $f$-graph, confirming this idea. Interestingly this graph is also the
first $f$-graph with a coefficient different from $\pm 1$ it having
the coefficient 2.

The integrands we find are given in a local form in configuration space, which is very closely related to the twistor integrands of \cite{nima1,nima2} as we demonstrated in Section 7 in the text: the twistor numerators involving parity odd parts can be rather painlessly rewritten in terms of simple squares of distances and the structure $\epsilon_{12345v}=\epsilon(X_1 X_2 X_3 X_4 X_5 X_v)$ where the X are coordinates on the projective light-cone in 6d related to those of Minkowski space, see appendix B. This object is conformally invariant and can be broken down to a sum of 4d terms of the type $x_{1v}^2 \epsilon(x_{2v} x_{3v} x_{4v} x_{5v})$. In the 6d epsilon, 1,2,3,4,5 denote the outer points, and only the sixth variable is an integration point.
All parity odd terms in our results are of this type; epsilon terms with more than one integration vertex do not occur. By the use of Schouten identities etc. one can remove any given point from an epsilon contraction, but at the expense of introducing further denominator factors. Hence there is freedom as to the writing of the end result, although the form we found is perhaps the most natural one since it is manifestly free of higher poles like $1/x_{ij}^4$.

Interestingly, it is possible to generate the parity even part of the
five-point amplitude from the parity odd bit up to four-loops 
using a few universal rules for how to replace an epsilon term. These
rules depend on the other numerator terms multiplying the $\epsilon_{12345v}$.
For example 
clearly the one-loop result can be rewritten as a single pentagon upon replacing
\begin{align}
i \epsilon_{123456} \ \rightarrow  \ \Big(
 x^2_{16} x^2_{24} x^2_{35} + 
 x^2_{26} x^2_{14} x^2_{35}+ 
x^2_{36} x^2_{14} x^2_{25}  + x^2_{46} x^2_{13} x^2_{25} + x^2_{56} x^2_{13} x^2_{24}
   + i  \epsilon_{123456}\Big)\ .
\end{align}
This is the only parity odd graph with a numerator involving an
$\epsilon$ and nothing else. Other numerators have various $x^2$
products multiplying. If we make the following replacements for $a,b,c>0$:
\begin{align}
  \label{eq:5}
 i  x_{13}^{2a} \epsilon_{123456} \ &\rightarrow  \ x_{13}^{2a}\Big(
 x^2_{56} x^2_{13} x^2_{24} + 
 x^2_{46} x^2_{13} x^2_{25} +
 x^2_{26} x^2_{14} x^2_{35}  + i  \epsilon_{123456}\Big)\notag\\
  i  x_{13}^{2a} x_{14}^{2b} \epsilon_{123456} \ &\rightarrow  \ x_{13}^{2a}x_{14}^{2b}\Big(
 x^2_{56} x^2_{13} x^2_{24}  +
 x^2_{26} x^2_{14} x^2_{35}  - 
 x^2_{16} x^2_{23} x^2_{45}+ i  \epsilon_{123456}\Big)\notag\\
 i  x_{13}^{2a} x_{24}^{2b} \epsilon_{123456} \ &\rightarrow  \ x_{13}^{2a}x_{24}^{2b}\Big(
 x^2_{56} x^2_{13} x^2_{24}  +
 x^2_{26} x^2_{14} x^2_{35}  + 
 x^2_{36} x^2_{14} x^2_{25}+ i  \epsilon_{123456}\Big)\notag\\
 i  x_{13}^{2a} x_{14}^{2b}x_{24}^{2c} \epsilon_{123456} \ &\rightarrow  \ x_{13}^{2a}x_{14}^{2b}x_{24}^{2c}\Big(
 x^2_{56} x^2_{13} x^2_{24}  +
 x^2_{26} x^2_{14} x^2_{35}  +
 x^2_{36} x^2_{14} x^2_{25} \notag\\
& \qquad \qquad \qquad \qquad \qquad 
- 
 x^2_{16} x^2_{23} x^2_{45}-
 x^2_{46} x^2_{13} x^2_{25}+ i  \epsilon_{123456}\Big)
\end{align}
and all forms related by cyclicty related in a similar way, then the parity odd graphs will give the
parity even graphs for free up to four loops. Beyond one loop, the easiest case to
check is obviously the two loop case~(\ref{eq:22}) where we use
the first replacement.  
 This procedure
fails for the first time at 5 loops where we are left with a single parity even graph
which is not determined by the parity odd sector in this manner:
\begin{center}
  \includegraphics[width=2.5cm]{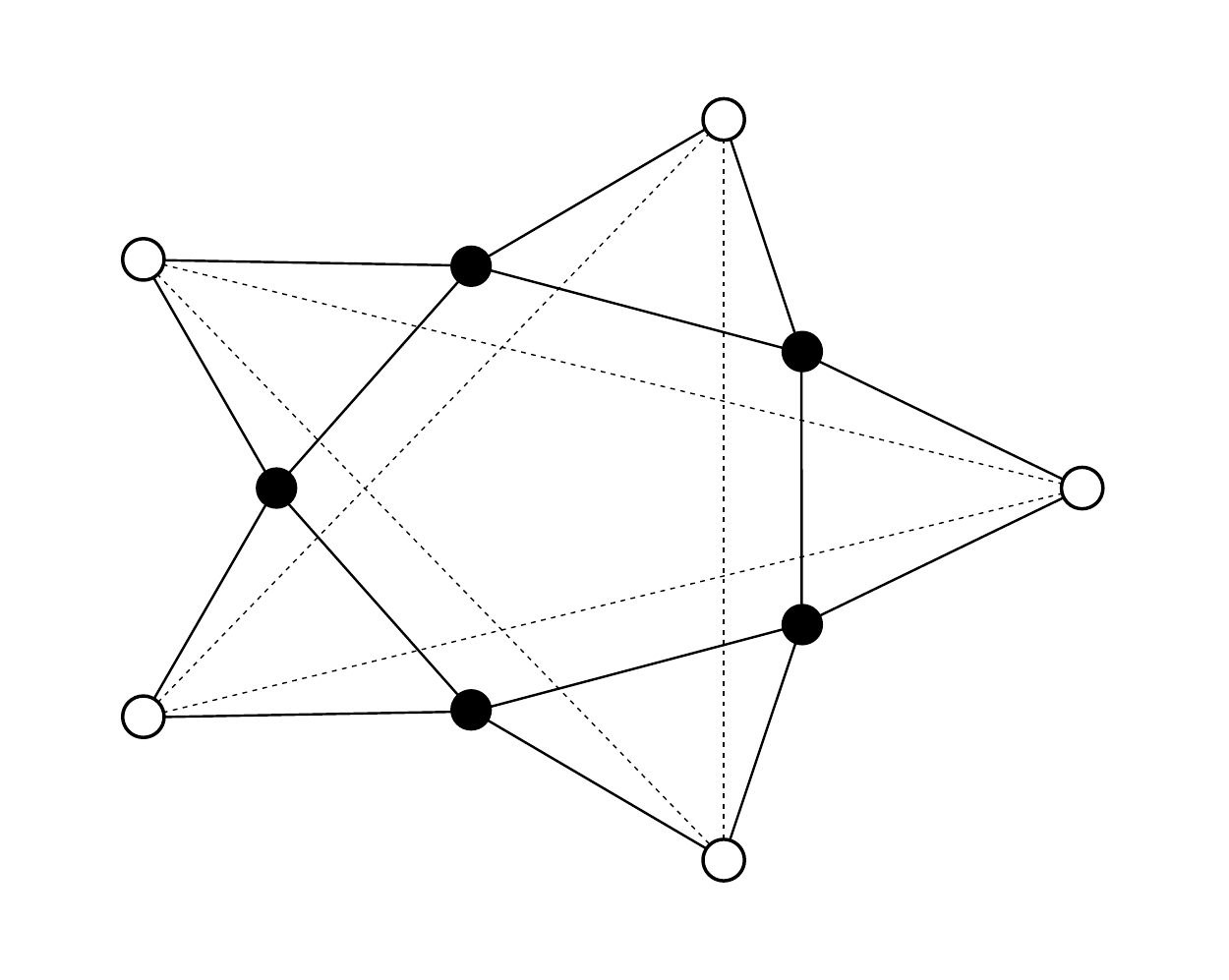}
\end{center}
This happens to be the single five-point amplitude graph generated by the
ten-point $f$-graph above whose coefficient is undetermined by
consistency with the duality. 
So we see that 
these rules for obtaining parity even graphs from parity odd are intimatey related to the consistency of the whole
system but we have not yet fully probed this.

Note that the twistor numerators of~\cite{nima1,nima2} (c.f. Section \ref{sec:twistor-integrands}) also combine even and
odd graphs and so the above rewriting may give expressions closer to
those. One direction for future work might indeed be to look for a
universal numerator describing higher-loop $n$-point amplitudes.

Another direction for future work would be to consider the six-point
light-like limit. Defining  
\begin{align}
F_6^{(\ell)}:=  \text{external factor} \times \lim_{\substack{x^2_{i\,i+1}\rightarrow 0 \\
    (\text{mod } 6)
}} 
 \int d^4x_7 \dots d^4x_{6+\ell} {f^{(\ell+2)} \over  \ell! }
\end{align}
where here the external factor is ${
  x^2_{12}x^2_{23}x^2_{34}x^2_{45}x^2_{56}x^2_{61}
  x^2_{13}x^2_{24}x^2_{35}x^2_{46}x^2_{51}x^2_{62}}$ then we will find the formula
\begin{align}
\sum_{\ell \geq 0} {a^\ell } {F_6^{(\ell)}}  \ =\ M_6 \overline{M}_6 +
\text{NMHV contribution}\ .
\end{align}

There are various complications arising here. Firstly the NMHV
contribution needs to be separated out (although this may be possible
due to singularities in $x_{14}^2, x_{25}^2$ and $x_{36}^2$  which can
only appear here and not in the MHV sector). Another complication
arises since there is no longer a distinction between product graphs
and disc planar graphs. The graph (one loop box)$\times$ (one loop box)
can appear in a disc planar fashion and indeed  does
appear in the two-loop six point result. Nevertheless we have seen
that one can obtain more information than appears at first sight from
these considerations and this certainly deserves further investigation.

We note that there are not believed to be any $\mu$ terms at
five-points and thus the results here should be valid to all orders in
dimensional regularisation parameter
$\epsilon$ by writing the integrals in momentum space and allowing the integration
momenta to live in $4-2 \, \epsilon$ dimensions.

Finally, our results at five loops and beyond are contained in various attachments to the electronic version of this article, as detailed at the end of Section 6.

\section*{Acknowledgements} 

We acknowledge support
from the EU Initial Training Network in High Energy Physics and
Mathematics: GATIS. BE is supported by DFG ``eigene Stelle'' Ed 78/4-2.
PH would like to acknowledge support from STFC  Consolidated
Grant number ST/J000426/1, TG gratefully acknowledges support from an EPSRC
studentship. RGA is supported by Mutua Madrile\~na foundation.

\appendix

\section*{Appendices}

\section{The operation ``cyc[]''}
It  is defined to be the weighted sum over cyclically
ordered (including parity flip) external
points.
Explicitly, for a function depending on only four out of the five external points this is defined as
\begin{align}
  \label{eq:18}
  \text{cyc} \left[f(x_1,x_2,x_3,x_4)\right]:= \sum_{1\leq
      i<j<k<l<i+5\leq 10}{
    f(x_i,x_j,x_k,x_l)+ f(x_l,x_k,x_j,x_i) \over \text{symmetry factor}[f]}
\end{align}
whereas for a 5-point function it is defined as
\begin{equation}
\begin{split}
  \label{eq:18b}
  \text{cyc} \left[f(x_1,x_2,x_3,x_4,x_5)\right]&:= \sum_{1\leq
      i<j<k<l<m<i+5\leq 10}{
    f(x_i,x_j,x_k,x_l,x_m)+ f(x_m,x_l,x_k,x_j,x_i) \over \text{symmetry
      factor}[f]} \\
&= \sum_{i=1}^5
  {f(x_{i},x_{i+1},x_{i+2},x_{i+3},x_{i+4})+
    f(x_{i+4},x_{i+3},x_{i+2},x_{i+1},x_{i}) \over \text{symmetry
      factor}[f]}\ .
\end{split}
\end{equation}
Here all points are external and are mod 5.
The symmetry factor is defined as the number of terms left invariant
under such permutations. So for example, for a four-point function
\begin{align}
  \label{eq:19}
  \text{symmetry
    factor}\left[f(x_1,x_2,x_3,x_4)\right]=&\Big|\big\{f(x_i,x_j,x_k,x_l)=f(x_1,x_2,x_3,x_4):
      {\scriptstyle 1\leq
      i<j<k<l<i+5\leq 10} \big\}\Big| \notag\\ 
&+ \Big|\big\{f(x_l,x_k,x_j,x_i)=f(x_1,x_2,x_3,x_4): {\scriptstyle 1\leq
      i<j<k<l<i+5\leq 10 } \big\} \Big|
\end{align}
and similarly for a five-point function. Note that this insures that
the argument of the operation cyc$[]$ always appears with weight 1
when expanding the result into inequivalent terms, i.e.
\begin{align}
  \label{eq:20b}
  \text{cyc} [f(x_1,x_2,x_3,x_4)] = f(x_1,x_2,x_3,x_4) + \dots 
\end{align}
Finally we note that we will be dealing with integrands in general. We define integrands
to be equal if they are equal up to
a permutation of internal points, i.e. our functions have hidden
dependence on internal variables
$f(x_1,x_2,x_3,x_4,x_5)=f(x_1,x_2,x_3,x_4,x_5;x_6, \dots x_{5+\ell})$
and we say
\begin{align}
  f(x_1,x_2,x_3,x_4,x_5 )&=  f(x_i,x_j,x_k,x_l,x_m ) \notag \\
\label{eq:21}  
&\Updownarrow\\
  f(x_1,x_2,x_3,x_4,x_5;x_6,x_7,\dots x_{5+\ell} )&=
  f(x_i,x_j,x_k,x_l,x_m;x_{\sigma_6},x_{\sigma_7},\dots
  x_{\sigma_{5+\ell}} )\notag
\end{align}
for some permutation $\sigma$ of the internal variables $x_6, \dots x_{5+\ell}$.

\section{4d Minkowski coordinates in 6d $X$-variables}
\label{sec:4d-mink-coord}

In order to relate our five-point integrands to similar twistor integrands found in the literature and also to explain the origin of $\epsilon_{12345v}$ used for our parity odd integrands, it is extremely useful to  view 4-dimensional Minkowski space as the (Klein) quadric inside $\mathbb{RP}^{5}$. This, and its relation to momentum twistors in the context of dual conformal symmetry for amplitudes was introduced in~\cite{0909.0250}.

Specifically we can describe Minkowski space in terms of six projective coordinates $X_{I}$ living in 2+4 dimensions and satisfying the null condition
\begin{equation}
X_{-1}^{2}+X_{0}^{2}-X_{1}^{2}-X_{2}^{2}-X_{3}^{2}-X_{4}^{2}=0
\end{equation}
As such the conformal group SO(2,4) then acts linearly on these coordinates. The four-dimensional Minkowski space coordinates $x^{\mu}$, $\mu=0,1,2,3$ can be obtained easily from these by choosing a suitable representation for the homogeneous coordinate $X^{I}$.
\begin{equation}\label{eq:32}
X^{I} \sim \left( \frac{1-x^{2}}{2}, x^{\mu}, \frac{1+x^{2}}{2} \right) \hspace{2cm} I=-1,0,1,2,3,4
\end{equation}

It is also useful --- especially in relation to twistor integrands ---
to consider the spinorial representation of the $X^I$. Using that $SO(2,4) \sim SU(2,2)$ it is this representation which we employ later to consider the integrands. There are two versions:
\begin{align}
X^{I} \rightarrow X=(\Sigma_{I})X^{I} \nonumber \\
X^{I} \rightarrow \tilde{X} = (\tilde{\Sigma}_{I})X^{I}
\end{align}
where  the $\Sigma$'s are are $4 \times 4$ sigma matrices in 6 dimensions.
We can choose them to satisfy
\begin{equation}
(\tilde{\Sigma}_{I})_{AB} = \frac{1}{2} \epsilon_{ABCD}( \Sigma_{I})^{CD}
\end{equation}
giving the relation 
\begin{equation} 
\tilde{X}_{AB} = \frac{1}{2}\epsilon_{ABCD} X^{CD} .\label{parityflip}
\end{equation}
The Clifford algebra relations,
\begin{equation}
 \Sigma_{I} \tilde{\Sigma}_{J} + \Sigma_{J} \tilde{\Sigma}_{I} = \tilde{\Sigma}_{I} \Sigma_{J} + \Sigma_{I} \tilde{\Sigma}_{J} = 2 \, \eta_{IJ}
\end{equation}
where $\eta_{IJ}$ is the flat metric in 2+4 dimensions, imply the following for any 6-vectors X, Y
\begin{align}
X \tilde{Y} + Y \tilde{X} = 2 \, X \cdot Y & \hspace{2cm} X \tilde{X} = X
\cdot X\ .
\end{align}
We also have that
\begin{align}
 \epsilon_{IJKLMN} \Sigma_{I} \tilde \Sigma_{J}\Sigma_{K}\tilde\Sigma_{L}\Sigma_{M}\tilde\Sigma_{N}= i 1_4\ .
\end{align}

\subsection{Different forms for $\epsilon_{123456}$}
\label{sec:diff-forms-epsil}
We now consider how we are to go about writing down conformal covariants. This can be done either using vector $X$'s or spinorial $X$'s, in both cases we simply need to soak up all indices. In the vectorial notation we can essentially use $\eta_{IJ}$ or $\epsilon_{IJKLMN}$ to form invariants, those obtained using a single $\epsilon_{IJKLMN}$ will be parity-odd. The covariants for 5-points and below must necessarily be composed of only $(X_{I} \cdot X_{J})$ whereas at six-points and above we may also have the parity-odd object
\begin{align}
 \epsilon_{123456}:= \epsilon_{IJKLMN} X^I_{1}X^J_{2}X^K_{3}X^L_{4}X^M_{5}X^N_{6}
=\det\Big(X_i{}^I\Big)\end{align}
 Indeed one can see that at six points this is the unique parity odd covariant piece. One can convert these invariants to four-dimensional notation straightforwardly by using~(\ref{eq:32})
\begin{align}
&X_{I} \cdot X_{J} \sim  -\frac12 (x_{i} - x_{j})^{2} \\
&\epsilon_{123456} \sim  \frac{1}{2} \frac1{4!} \sum_{\sigma \epsilon S_{6}} (-1)^{\sigma} x_{\sigma_1}^{2} \epsilon(x_{\sigma_2},x_{\sigma_3},x_{\sigma_4},x_{\sigma_5})
\end{align}
Note that the latter expression for $\epsilon_{123456}$ does not look translation invariant in terms of Minkowski space variables.  But of course it must be, and indeed an infinitesimal translation $x_i \rightarrow x_i + a$ gives
\begin{align}
  \delta \epsilon_{123456} =  \sum_{\sigma \epsilon S_{6}} (-1)^{\sigma} \Big(a\cdot x_{\sigma_1} \epsilon(x_{\sigma_2},x_{\sigma_3},x_{\sigma_4},x_{\sigma_5})+2 x_{\sigma_1}^2\epsilon(a,x_{\sigma_3},x_{\sigma_4},x_{\sigma_5})\Big)=0\ 
\end{align}
The first term vanishes due to a five term identity (expressing the fact that five points in $M^4$ are linearly dependent) and the second term vanishes since it is independent of $x_{\sigma_2}$ and $x_{\sigma_6}$ which appear antisymmetrically.

We can therefore rewrite the expression to make the translation invariance manifest, eg by translating by $x_6$ giving
\begin{align}
  \epsilon_{123456} &\sim  \frac{1}{2} \frac1{4!} \sum_{\sigma \epsilon S_{5}} (-1)^{\sigma} x_{\sigma_1\, 6}^{2} \epsilon(x_{\sigma_2\, 6},x_{\sigma_3\, 6},x_{\sigma_4\, 6},x_{\sigma_5\, 6})\ .
\end{align}

It is also useful however (to compare with results in the literature which are often written in terms of scalar integrals) to give another writing  of the object $\epsilon_{123456}$.

For this we first define the 6-vector $Y$ as
\begin{align}
  Y_I = \epsilon_{IJKLMN}X_1^J X_2^K X_3^L X_4^M X_5^N
\end{align}
so that $\epsilon_{123456}=-Y.X_6$ and then we decompose $Y$ in terms
of the five vectors $X_i$ and the  vector $I=(1,0,0,0,0,-1)$ which
represents infinity in Minkowski space. Including infinity breaks
conformal invariance but allows us to write all integrands in terms of
purely scalar integrands which is common  in the literature.

So we write
\begin{align}\label{eq:39}
  Y = \sum_{i=1}^5 \alpha_i X_i + \beta I
\end{align}
and if we can solve  for the coefficients $\alpha_i,\beta$ 
we thus have an expression for $\epsilon_{123456}$ as
\begin{align}
  \label{eq:38}
  \epsilon_{123456}= -\sum_{i=1}^5 \alpha_i X_i.X_6 - \beta =
  1/2\sum_{i=1}^5 \alpha_i x_{i6}^2 - \beta\ .
\end{align}

To solve for by the $\alpha_i,\beta$ simply dot~(\ref{eq:39}) with
$X_j, Y$, to obtain the matrix  equation
\begin{align}\label{eq:34}
\left(\left(
  \begin{array}{c}
X_i\\I
\end{array}\right)
.(X_j,I)\right)
\left(\begin{array}{c}
  \alpha_i\\\beta
\end{array}\right)
=
\left(\begin{array}{c}
  0\\Y.I
\end{array}\right)
\end{align}
where we use that $X_j.I= 1$. Noting further that 
\begin{align}
  Y.I = -\det\left(
    \begin{array}{c}
X_i{}^I\\ I^I
\end{array}
\right)     =   \frac1{4!}\sum_{i\in S_5} (-1)^{\sigma} \epsilon(x_{\sigma_1}, x_{\sigma_2}, x_{\sigma_3}, x_{\sigma_4}) = \epsilon(x_{12}, x_{23}, x_{34}, x_{45})\ ,
\end{align}
where the last equation can be obtained by simply expanding out the
right-hand side which will be recognised as
$\epsilon(p_1,p_2,p_3,p_4)$ in momentum space. Further useful equations
straightforward to derive are 
\begin{align}
  \label{eq:33}
  \det(X_i.X_j)&=-\frac1{16}x_{13}^2 x_{24}^2x_{35}^2x_{41}^2x_{52}^2\\
  \det\left(\left(
  \begin{array}{c}
X_i\\I
\end{array}\right)
.(X_j,I)\right)&=\det\left(
    \begin{array}{c}
X_i{}^I\\ I^I
\end{array}\right)^2 = \epsilon(x_{12},x_{23},x_{34},x_{45})^2\ .
\end{align}
(Note that the first equation above is the only one which is not 
valid for arbitrary values of space-time points $X_i$, but only in the five-point light-like limit where
$X_i.X_{i+1}=0$.): 
Thus inverting~(\ref{eq:34}) and using these formulae we obtain
\begin{align}
  \label{eq:35}
  \alpha_1 &=  \frac{-x^2_{24} x^2_{35}\left( x_{13}^2x_{52}^2 -
    x^2_{24}x^2_{13} + x^2_{35}x^2_{24} - x^2_{41}x^2_{35} + x^2_{52}x^2_{41}
    \right)}{16 \,\epsilon(x_{12},x_{23},x_{34},x_{45})}  \qquad  \text{and cyclic for }\alpha_2,\dots \alpha_5 \notag \\ \beta&=
    \frac{-x^2_{13} x^2_{24}
      x^2_{35}x^2_{41}x^2_{52}}{16\,\epsilon(x_{12},x_{23},x_{34},x_{45})}\ .
\end{align}
The $\alpha_i$ can be further rewritten in the simpler form in
momentum space via 
\begin{align}
  \label{eq:36}
\alpha_1 = \frac{s_2 s_3}{4 \,\epsilon(p_1,p_2,p_3,p_4)}
 Tr(\slashed{p}_4 \slashed{p}_5 \slashed{p}_1 \slashed{p}_2) \quad
 \text{and cyclic for }\alpha_2,\dots \alpha_5
\end{align}
where $p_i=x_{i\,i+1},\ s_i=x_{i\,i+2}^2 = (p_i+p_{i+1})^2$.

\end{document}

%% file: 1loopf.tex
\begin{aligned}
 \mathcal{I}_1^{(1)}&=\text{cyc}\left[\frac{x_{13}^2 x_{25}^2}{x_{16}^2 x_{26}^2 x_{36}^2 x_{56}^2}\right] & \mathcal{I}_2^{(1)}&=\text{cyc}\left[\frac{i \epsilon _{123456}}{x_{16}^2 x_{26}^2 x_{36}^2 x_{46}^2 x_{56}^2}\right] \\
\end{aligned}

%% file: 2loopf.tex
\begin{aligned}
 \mathcal{I}_1^{(2)}&=\text{cyc}\left[\frac{x_{13}^4 x_{25}^2}{x_{16}^2 x_{17}^2 x_{27}^2 x_{36}^2 x_{37}^2 x_{56}^2 x_{67}^2}\right] & \mathcal{I}_2^{(2)}&=\text{cyc}\left[\frac{x_{16}^2 x_{24}^2 x_{25}^2 x_{35}^2}{x_{17}^2 x_{26}^2 x_{27}^2 x_{36}^2 x_{46}^2 x_{56}^2 x_{57}^2 x_{67}^2}\right] \\
 \mathcal{I}_3^{(2)}&=\text{cyc}\left[\frac{i x_{13}^2 \epsilon _{123456}}{x_{16}^2 x_{17}^2 x_{27}^2 x_{36}^2 x_{37}^2 x_{46}^2 x_{56}^2 x_{67}^2}\right] & \text{} \\
\end{aligned}

%% file: 3loopf.tex
\begin{aligned}
 \mathcal{I}_1^{(3)}&=\text{cyc}\left(\frac{x_{13}^6 x_{25}^2}{x_{16}^2 x_{17}^2 x_{18}^2 x_{28}^2 x_{36}^2 x_{37}^2 x_{38}^2 x_{57}^2 x_{67}^2 x_{68}^2}\right) & \mathcal{I}_2^{(3)}&=\text{cyc}\left(\frac{x_{16}^4 x_{24}^2 x_{25}^2 x_{35}^2}{x_{17}^2 x_{18}^2 x_{26}^2 x_{28}^2 x_{36}^2 x_{46}^2 x_{56}^2 x_{57}^2 x_{67}^2 x_{68}^2 x_{78}^2}\right) \\
 \mathcal{I}_3^{(3)}&=\text{cyc}\left(\frac{x_{13}^4 x_{25}^2 x_{35}^2 x_{46}^2}{x_{16}^2 x_{18}^2 x_{28}^2 x_{36}^2 x_{37}^2 x_{38}^2 x_{47}^2 x_{56}^2 x_{57}^2 x_{67}^2 x_{68}^2}\right) & \mathcal{I}_4^{(3)}&=\text{cyc}\left(\frac{x_{13}^4 x_{24}^2 x_{46}^2}{x_{16}^2 x_{18}^2 x_{26}^2 x_{36}^2 x_{37}^2 x_{47}^2 x_{48}^2 x_{67}^2 x_{68}^2 x_{78}^2}\right) \\
 \mathcal{I}_5^{(3)}&=\text{cyc}\left(\frac{x_{14}^2 x_{16}^2 x_{24}^2 x_{25}^2 x_{37}^2}{x_{17}^2 x_{18}^2 x_{26}^2 x_{28}^2 x_{36}^2 x_{46}^2 x_{47}^2 x_{57}^2 x_{67}^2 x_{68}^2 x_{78}^2}\right) & \mathcal{I}_6^{(3)}&=\text{cyc}\left(\frac{x_{16}^2 x_{24}^2 x_{25}^4 x_{35}^2}{x_{18}^2 x_{26}^2 x_{27}^2 x_{28}^2 x_{36}^2 x_{46}^2 x_{56}^2 x_{57}^2 x_{58}^2 x_{67}^2 x_{78}^2}\right) \\
 \mathcal{I}_7^{(3)}&=\text{cyc}\left(\frac{x_{13}^2 x_{24}^2 x_{25}^2 x_{35}^2}{x_{18}^2 x_{26}^2 x_{28}^2 x_{36}^2 x_{37}^2 x_{47}^2 x_{57}^2 x_{58}^2 x_{67}^2 x_{68}^2}\right) & \mathcal{I}_8^{(3)}&=\text{cyc}\left(\frac{x_{13}^2 x_{14}^2 x_{35}^2}{x_{16}^2 x_{17}^2 x_{36}^2 x_{38}^2 x_{48}^2 x_{57}^2 x_{67}^2 x_{68}^2 x_{78}^2}\right) \\
 \mathcal{I}_9^{(3)}&=\text{cyc}\left(\frac{i x_{13}^4 \epsilon _{123456}}{x_{16}^2 x_{17}^2 x_{18}^2 x_{28}^2 x_{36}^2 x_{37}^2 x_{38}^2 x_{46}^2 x_{56}^2 x_{67}^2 x_{78}^2}\right) & \mathcal{I}_{10}^{(3)}&=\text{cyc}\left(\frac{i x_{13}^2 x_{14}^2 \epsilon _{123456}}{x_{16}^2 x_{17}^2 x_{18}^2 x_{28}^2 x_{36}^2 x_{38}^2 x_{46}^2 x_{47}^2 x_{57}^2 x_{67}^2 x_{68}^2}\right) \\
 \mathcal{I}_{11}^{(3)}&=\text{cyc}\left(\frac{i x_{24}^2 x_{36}^2 \epsilon _{123456}}{x_{16}^2 x_{26}^2 x_{28}^2 x_{37}^2 x_{38}^2 x_{46}^2 x_{47}^2 x_{56}^2 x_{67}^2 x_{68}^2 x_{78}^2}\right) & \mathcal{I}_{12}^{(3)}&=\text{cyc}\left(\frac{i x_{14}^2 x_{27}^2 \epsilon _{123456}}{x_{17}^2 x_{18}^2 x_{26}^2 x_{28}^2 x_{36}^2 x_{46}^2 x_{47}^2 x_{57}^2 x_{67}^2 x_{68}^2 x_{78}^2}\right) \\
 \mathcal{I}_{13}^{(3)}&=\text{cyc}\left(\frac{i x_{13}^2 \epsilon _{123456}}{x_{16}^2 x_{17}^2 x_{26}^2 x_{36}^2 x_{38}^2 x_{48}^2 x_{57}^2 x_{67}^2 x_{68}^2 x_{78}^2}\right) & \text{} \\
\end{aligned}

%% file: 4loopf.tex
 \mathcal{I}_1^{(4)}&=\text{cyc}\left[\frac{x_{13}^8 x_{25}^2}{x_{16}^2 x_{17}^2 x_{18}^2 x_{19}^2 x_{29}^2 x_{36}^2 x_{37}^2 x_{38}^2 x_{39}^2 x_{57}^2 x_{68}^2 x_{69}^2 x_{78}^2}\right] & \mathcal{I}_2^{(4)}&=\text{cyc}\left[\frac{x_{16}^6 x_{24}^2 x_{25}^2 x_{35}^2}{x_{17}^2 x_{18}^2 x_{19}^2 x_{26}^2 x_{29}^2 x_{36}^2 x_{46}^2 x_{56}^2 x_{58}^2 x_{67}^2 x_{68}^2 x_{69}^2 x_{78}^2 x_{79}^2}\right] \\
 \mathcal{I}_3^{(4)}&=\text{cyc}\left[\frac{x_{13}^2 x_{24}^2 x_{25}^2 x_{35}^2}{x_{19}^2 x_{28}^2 x_{29}^2 x_{37}^2 x_{38}^2 x_{47}^2 x_{56}^2 x_{59}^2 x_{67}^2 x_{68}^2 x_{69}^2 x_{78}^2}\right] & \mathcal{I}_4^{(4)}&=\text{cyc}\left[\frac{x_{14}^2 x_{24}^2 x_{25}^2}{x_{17}^2 x_{26}^2 x_{29}^2 x_{48}^2 x_{49}^2 x_{57}^2 x_{67}^2 x_{68}^2 x_{69}^2 x_{78}^2 x_{89}^2}\right] \\
 \mathcal{I}_5^{(4)}&=\text{cyc}\left[\frac{x_{13}^6 x_{25}^2 x_{35}^2 x_{46}^2}{x_{16}^2 x_{17}^2 x_{19}^2 x_{29}^2 x_{36}^2 x_{37}^2 x_{38}^2 x_{39}^2 x_{48}^2 x_{56}^2 x_{58}^2 x_{67}^2 x_{68}^2 x_{79}^2}\right] & \mathcal{I}_6^{(4)}&=\text{cyc}\left[\frac{x_{13}^6 x_{24}^2 x_{67}^2}{x_{16}^2 x_{17}^2 x_{19}^2 x_{26}^2 x_{36}^2 x_{37}^2 x_{38}^2 x_{47}^2 x_{68}^2 x_{69}^2 x_{78}^2 x_{79}^2 x_{89}^2}\right] \\
 \mathcal{I}_7^{(4)}&=\text{cyc}\left[\frac{x_{14}^6 x_{35}^2 x_{36}^2}{x_{16}^2 x_{17}^2 x_{19}^2 x_{38}^2 x_{39}^2 x_{46}^2 x_{47}^2 x_{48}^2 x_{57}^2 x_{67}^2 x_{68}^2 x_{69}^2 x_{89}^2}\right] & \mathcal{I}_8^{(4)}&=\text{cyc}\left[\frac{x_{14}^2 x_{16}^4 x_{24}^2 x_{25}^2 x_{37}^2}{x_{17}^2 x_{18}^2 x_{19}^2 x_{26}^2 x_{29}^2 x_{36}^2 x_{46}^2 x_{47}^2 x_{57}^2 x_{67}^2 x_{68}^2 x_{69}^2 x_{78}^2 x_{89}^2}\right] \\
 \mathcal{I}_9^{(4)}&=\text{cyc}\left[\frac{x_{16}^4 x_{24}^2 x_{25}^2 x_{35}^4}{x_{17}^2 x_{19}^2 x_{26}^2 x_{29}^2 x_{36}^2 x_{38}^2 x_{48}^2 x_{56}^2 x_{57}^2 x_{58}^2 x_{67}^2 x_{68}^2 x_{69}^2 x_{79}^2}\right] & \mathcal{I}_{10}^{(4)}&=\text{cyc}\left[\frac{x_{16}^4 x_{24}^2 x_{25}^2 x_{35}^2 x_{57}^2}{x_{17}^2 x_{19}^2 x_{26}^2 x_{27}^2 x_{36}^2 x_{46}^2 x_{56}^2 x_{58}^2 x_{59}^2 x_{67}^2 x_{68}^2 x_{78}^2 x_{79}^2 x_{89}^2}\right] \\
 \mathcal{I}_{11}^{(4)}&=\text{cyc}\left[\frac{x_{13}^6 x_{14}^2 x_{27}^2 x_{35}^2}{x_{16}^2 x_{17}^2 x_{18}^2 x_{19}^2 x_{26}^2 x_{36}^2 x_{37}^2 x_{38}^2 x_{39}^2 x_{47}^2 x_{57}^2 x_{69}^2 x_{78}^2 x_{89}^2}\right] & \mathcal{I}_{12}^{(4)}&=\text{cyc}\left[\frac{x_{13}^2 x_{17}^4 x_{24}^4}{x_{16}^2 x_{18}^2 x_{19}^2 x_{26}^2 x_{27}^2 x_{37}^2 x_{47}^2 x_{48}^2 x_{67}^2 x_{69}^2 x_{78}^2 x_{79}^2 x_{89}^2}\right] \\
 \mathcal{I}_{13}^{(4)}&=\text{cyc}\left[\frac{x_{16}^4 x_{24}^2 x_{25}^2 x_{35}^2 x_{37}^2}{x_{17}^2 x_{18}^2 x_{27}^2 x_{29}^2 x_{36}^2 x_{39}^2 x_{46}^2 x_{56}^2 x_{58}^2 x_{67}^2 x_{68}^2 x_{69}^2 x_{78}^2 x_{79}^2}\right] & \mathcal{I}_{14}^{(4)}&=\text{cyc}\left[\frac{x_{13}^4 x_{17}^2 x_{25}^2 x_{35}^2 x_{46}^2}{x_{16}^2 x_{18}^2 x_{19}^2 x_{29}^2 x_{36}^2 x_{37}^2 x_{39}^2 x_{47}^2 x_{57}^2 x_{58}^2 x_{67}^2 x_{68}^2 x_{69}^2 x_{78}^2}\right] \\
 \mathcal{I}_{15}^{(4)}&=\text{cyc}\left[\frac{x_{13}^4 x_{17}^2 x_{24}^2 x_{46}^2}{x_{16}^2 x_{18}^2 x_{19}^2 x_{26}^2 x_{36}^2 x_{37}^2 x_{47}^2 x_{48}^2 x_{67}^2 x_{69}^2 x_{78}^2 x_{79}^2 x_{89}^2}\right] & \mathcal{I}_{16}^{(4)}&=\text{cyc}\left[\frac{x_{13}^4 x_{14}^4 x_{24}^2 x_{56}^2}{x_{16}^2 x_{17}^2 x_{18}^2 x_{19}^2 x_{29}^2 x_{36}^2 x_{39}^2 x_{46}^2 x_{47}^2 x_{48}^2 x_{58}^2 x_{67}^2 x_{69}^2 x_{78}^2}\right] \\
 \mathcal{I}_{17}^{(4)}&=\text{cyc}\left[\frac{x_{16}^2 x_{18}^2 x_{24}^2 x_{25}^2 x_{35}^2 x_{37}^2}{x_{17}^2 x_{19}^2 x_{26}^2 x_{29}^2 x_{36}^2 x_{38}^2 x_{48}^2 x_{57}^2 x_{58}^2 x_{67}^2 x_{68}^2 x_{69}^2 x_{78}^2 x_{79}^2}\right] & \mathcal{I}_{18}^{(4)}&=\text{cyc}\left[\frac{x_{16}^2 x_{17}^2 x_{24}^2 x_{25}^4 x_{35}^2}{x_{18}^2 x_{19}^2 x_{26}^2 x_{27}^2 x_{29}^2 x_{36}^2 x_{46}^2 x_{56}^2 x_{57}^2 x_{58}^2 x_{67}^2 x_{78}^2 x_{79}^2 x_{89}^2}\right] \\
 \mathcal{I}_{19}^{(4)}&=\text{cyc}\left[\frac{x_{13}^2 x_{14}^2 x_{26}^2 x_{35}^4 x_{46}^2}{x_{16}^2 x_{19}^2 x_{29}^2 x_{36}^2 x_{37}^2 x_{39}^2 x_{47}^2 x_{48}^2 x_{56}^2 x_{58}^2 x_{67}^2 x_{68}^2 x_{69}^2 x_{78}^2}\right] & \mathcal{I}_{20}^{(4)}&=\text{cyc}\left[\frac{x_{13}^2 x_{14}^2 x_{26}^2 x_{35}^2 x_{36}^2 x_{47}^2}{x_{16}^2 x_{17}^2 x_{27}^2 x_{37}^2 x_{38}^2 x_{39}^2 x_{46}^2 x_{48}^2 x_{56}^2 x_{67}^2 x_{68}^2 x_{69}^2 x_{79}^2 x_{89}^2}\right] \\
 \mathcal{I}_{21}^{(4)}&=\text{cyc}\left[\frac{x_{13}^2 x_{14}^2 x_{24}^2 x_{26}^2 x_{36}^2 x_{57}^2}{x_{16}^2 x_{19}^2 x_{27}^2 x_{29}^2 x_{37}^2 x_{38}^2 x_{46}^2 x_{48}^2 x_{56}^2 x_{67}^2 x_{68}^2 x_{69}^2 x_{78}^2 x_{79}^2}\right] & \mathcal{I}_{22}^{(4)}&=\text{cyc}\left[\frac{x_{14}^2 x_{24}^4 x_{25}^2 x_{36}^2 x_{67}^2}{x_{16}^2 x_{26}^2 x_{27}^2 x_{29}^2 x_{37}^2 x_{46}^2 x_{47}^2 x_{48}^2 x_{56}^2 x_{68}^2 x_{69}^2 x_{78}^2 x_{79}^2 x_{89}^2}\right] \\
 \mathcal{I}_{23}^{(4)}&=\text{cyc}\left[\frac{x_{17}^2 x_{24}^2 x_{25}^4 x_{35}^2 x_{36}^2}{x_{19}^2 x_{26}^2 x_{28}^2 x_{29}^2 x_{37}^2 x_{38}^2 x_{47}^2 x_{56}^2 x_{57}^2 x_{59}^2 x_{67}^2 x_{68}^2 x_{69}^2 x_{78}^2}\right] & \mathcal{I}_{24}^{(4)}&=\text{cyc}\left[\frac{x_{14}^4 x_{24}^2 x_{25}^2 x_{27}^2 x_{36}^2}{x_{17}^2 x_{19}^2 x_{26}^2 x_{28}^2 x_{29}^2 x_{38}^2 x_{46}^2 x_{47}^2 x_{48}^2 x_{57}^2 x_{67}^2 x_{68}^2 x_{69}^2 x_{79}^2}\right] \\
 \mathcal{I}_{25}^{(4)}&=\text{cyc}\left[\frac{x_{13}^2 x_{18}^2 x_{24}^4 x_{36}^2}{x_{16}^2 x_{17}^2 x_{26}^2 x_{29}^2 x_{38}^2 x_{39}^2 x_{47}^2 x_{48}^2 x_{67}^2 x_{68}^2 x_{69}^2 x_{78}^2 x_{89}^2}\right] & \mathcal{I}_{26}^{(4)}&=\text{cyc}\left[\frac{x_{13}^2 x_{18}^2 x_{25}^2 x_{27}^2 x_{35}^2 x_{46}^2}{x_{17}^2 x_{19}^2 x_{26}^2 x_{29}^2 x_{36}^2 x_{38}^2 x_{48}^2 x_{57}^2 x_{58}^2 x_{67}^2 x_{68}^2 x_{69}^2 x_{78}^2 x_{79}^2}\right] \\
 \mathcal{I}_{27}^{(4)}&=\text{cyc}\left[\frac{x_{13}^2 x_{14}^2 x_{19}^2 x_{28}^2 x_{35}^2 x_{46}^2}{x_{16}^2 x_{17}^2 x_{18}^2 x_{26}^2 x_{36}^2 x_{39}^2 x_{48}^2 x_{49}^2 x_{58}^2 x_{67}^2 x_{69}^2 x_{78}^2 x_{79}^2 x_{89}^2}\right] & \mathcal{I}_{28}^{(4)}&=\text{cyc}\left[\frac{x_{18}^2 x_{24}^4 x_{25}^2 x_{35}^2 x_{67}^2}{x_{16}^2 x_{26}^2 x_{27}^2 x_{29}^2 x_{37}^2 x_{47}^2 x_{48}^2 x_{56}^2 x_{58}^2 x_{68}^2 x_{69}^2 x_{78}^2 x_{79}^2 x_{89}^2}\right] \\
 \mathcal{I}_{29}^{(4)}&=\text{cyc}\left[\frac{x_{13}^4 x_{25}^2 x_{35}^2}{x_{18}^2 x_{19}^2 x_{29}^2 x_{36}^2 x_{37}^2 x_{39}^2 x_{57}^2 x_{58}^2 x_{67}^2 x_{68}^2 x_{69}^2 x_{78}^2}\right] & \mathcal{I}_{30}^{(4)}&=\text{cyc}\left[\frac{x_{14}^4 x_{35}^2}{x_{18}^2 x_{19}^2 x_{39}^2 x_{46}^2 x_{47}^2 x_{56}^2 x_{67}^2 x_{68}^2 x_{78}^2 x_{79}^2 x_{89}^2}\right] \\
 \mathcal{I}_{31}^{(4)}&=\text{cyc}\left[\frac{x_{16}^2 x_{24}^2 x_{25}^2 x_{35}^2}{x_{19}^2 x_{26}^2 x_{29}^2 x_{36}^2 x_{48}^2 x_{57}^2 x_{58}^2 x_{67}^2 x_{68}^2 x_{69}^2 x_{78}^2 x_{79}^2}\right] & \mathcal{I}_{32}^{(4)}&=\text{cyc}\left[\frac{x_{14}^2 x_{16}^2 x_{24}^2 x_{25}^2 x_{35}^2}{x_{17}^2 x_{19}^2 x_{26}^2 x_{27}^2 x_{36}^2 x_{46}^2 x_{48}^2 x_{58}^2 x_{59}^2 x_{67}^2 x_{68}^2 x_{79}^2 x_{89}^2}\right] \\
 \mathcal{I}_{33}^{(4)}&=\text{cyc}\left[\frac{x_{13}^4 x_{14}^2 x_{24}^2 x_{35}^2}{x_{16}^2 x_{18}^2 x_{19}^2 x_{26}^2 x_{36}^2 x_{37}^2 x_{39}^2 x_{47}^2 x_{48}^2 x_{58}^2 x_{69}^2 x_{78}^2 x_{79}^2}\right] & \mathcal{I}_{34}^{(4)}&=\text{cyc}\left[\frac{x_{14}^2 x_{17}^2 x_{24}^2 x_{25}^2}{x_{16}^2 x_{19}^2 x_{26}^2 x_{27}^2 x_{47}^2 x_{48}^2 x_{58}^2 x_{67}^2 x_{69}^2 x_{78}^2 x_{79}^2 x_{89}^2}\right] \\
 \mathcal{I}_{35}^{(4)}&=\text{cyc}\left[\frac{x_{13}^2 x_{14}^2 x_{17}^2 x_{35}^2}{x_{16}^2 x_{18}^2 x_{19}^2 x_{37}^2 x_{39}^2 x_{47}^2 x_{58}^2 x_{67}^2 x_{68}^2 x_{69}^2 x_{78}^2 x_{79}^2}\right] & \mathcal{I}_{36}^{(4)}&=\text{cyc}\left[\frac{x_{13}^2 x_{14}^2 x_{17}^2 x_{25}^2 x_{35}^2}{x_{16}^2 x_{18}^2 x_{19}^2 x_{29}^2 x_{37}^2 x_{39}^2 x_{47}^2 x_{56}^2 x_{57}^2 x_{67}^2 x_{68}^2 x_{78}^2 x_{89}^2}\right] \\
 \mathcal{I}_{37}^{(4)}&=\text{cyc}\left[\frac{x_{16}^2 x_{24}^2 x_{25}^2 x_{35}^2}{x_{19}^2 x_{28}^2 x_{29}^2 x_{36}^2 x_{46}^2 x_{56}^2 x_{57}^2 x_{67}^2 x_{68}^2 x_{78}^2 x_{79}^2 x_{89}^2}\right] & \mathcal{I}_{38}^{(4)}&=\text{cyc}\left[\frac{x_{14}^2 x_{17}^2 x_{24}^2 x_{25}^2 x_{35}^2}{x_{16}^2 x_{18}^2 x_{28}^2 x_{29}^2 x_{39}^2 x_{47}^2 x_{49}^2 x_{56}^2 x_{57}^2 x_{67}^2 x_{68}^2 x_{78}^2 x_{79}^2}\right] \\
 \mathcal{I}_{39}^{(4)}&=\text{cyc}\left[\frac{x_{13}^4 x_{25}^4}{x_{16}^2 x_{19}^2 x_{27}^2 x_{29}^2 x_{37}^2 x_{38}^2 x_{56}^2 x_{58}^2 x_{68}^2 x_{69}^2 x_{78}^2 x_{79}^2}\right] & \mathcal{I}_{40}^{(4)}&=\text{cyc}\left[\frac{x_{13}^4 x_{14}^2 x_{27}^2 x_{35}^2}{x_{16}^2 x_{17}^2 x_{18}^2 x_{26}^2 x_{36}^2 x_{37}^2 x_{39}^2 x_{49}^2 x_{58}^2 x_{67}^2 x_{78}^2 x_{79}^2 x_{89}^2}\right] \\
 \mathcal{I}_{41}^{(4)}&=\text{cyc}\left[\frac{x_{16}^2 x_{24}^4 x_{25}^2 x_{35}^2}{x_{19}^2 x_{26}^2 x_{27}^2 x_{29}^2 x_{36}^2 x_{46}^2 x_{48}^2 x_{58}^2 x_{59}^2 x_{67}^2 x_{68}^2 x_{78}^2 x_{79}^2}\right] & \mathcal{I}_{42}^{(4)}&=\text{cyc}\left[\frac{x_{13}^2 x_{16}^2 x_{25}^2 x_{28}^2 x_{35}^2}{x_{17}^2 x_{18}^2 x_{26}^2 x_{27}^2 x_{36}^2 x_{39}^2 x_{58}^2 x_{59}^2 x_{67}^2 x_{68}^2 x_{69}^2 x_{78}^2 x_{89}^2}\right] \\
 \mathcal{I}_{43}^{(4)}&=\text{cyc}\left[\frac{x_{14}^2 x_{16}^2 x_{24}^2 x_{25}^2 x_{37}^2}{x_{17}^2 x_{18}^2 x_{26}^2 x_{27}^2 x_{36}^2 x_{46}^2 x_{49}^2 x_{58}^2 x_{67}^2 x_{69}^2 x_{78}^2 x_{79}^2 x_{89}^2}\right] & \mathcal{I}_{44}^{(4)}&=\text{cyc}\left[\frac{x_{13}^2 x_{14}^2 x_{17}^2 x_{24}^2 x_{35}^2 x_{69}^2}{x_{16}^2 x_{18}^2 x_{19}^2 x_{29}^2 x_{37}^2 x_{39}^2 x_{46}^2 x_{47}^2 x_{56}^2 x_{67}^2 x_{68}^2 x_{78}^2 x_{79}^2 x_{89}^2}\right] \\
 \mathcal{I}_{45}^{(4)}&=\text{cyc}\left[\frac{i x_{13}^6 \epsilon _{123456}}{x_{16}^2 x_{17}^2 x_{18}^2 x_{19}^2 x_{29}^2 x_{36}^2 x_{37}^2 x_{38}^2 x_{39}^2 x_{46}^2 x_{56}^2 x_{68}^2 x_{78}^2 x_{79}^2}\right] & \mathcal{I}_{46}^{(4)}&=\text{cyc}\left[\frac{i x_{13}^2 \epsilon _{123456}}{x_{16}^2 x_{19}^2 x_{26}^2 x_{36}^2 x_{38}^2 x_{47}^2 x_{57}^2 x_{68}^2 x_{69}^2 x_{78}^2 x_{79}^2 x_{89}^2}\right] \\
 \mathcal{I}_{47}^{(4)}&=\text{cyc}\left[\frac{i x_{13}^4 x_{14}^2 \epsilon _{123456}}{x_{16}^2 x_{17}^2 x_{18}^2 x_{19}^2 x_{29}^2 x_{36}^2 x_{37}^2 x_{39}^2 x_{46}^2 x_{48}^2 x_{58}^2 x_{67}^2 x_{68}^2 x_{79}^2}\right] & \mathcal{I}_{48}^{(4)}&=\text{cyc}\left[\frac{i x_{16}^4 x_{25}^2 \epsilon _{123456}}{x_{17}^2 x_{18}^2 x_{19}^2 x_{26}^2 x_{27}^2 x_{36}^2 x_{46}^2 x_{56}^2 x_{58}^2 x_{67}^2 x_{68}^2 x_{69}^2 x_{79}^2 x_{89}^2}\right] \\
 \mathcal{I}_{49}^{(4)}&=\text{cyc}\left[\frac{i x_{13}^4 x_{26}^2 \epsilon _{123457}}{x_{16}^2 x_{17}^2 x_{18}^2 x_{28}^2 x_{29}^2 x_{36}^2 x_{37}^2 x_{39}^2 x_{47}^2 x_{57}^2 x_{67}^2 x_{68}^2 x_{69}^2 x_{89}^2}\right] & \mathcal{I}_{50}^{(4)}&=\text{cyc}\left[\frac{i x_{13}^4 x_{46}^2 \epsilon _{123457}}{x_{16}^2 x_{17}^2 x_{19}^2 x_{29}^2 x_{36}^2 x_{38}^2 x_{39}^2 x_{47}^2 x_{48}^2 x_{57}^2 x_{67}^2 x_{68}^2 x_{69}^2 x_{78}^2}\right] \\
 \mathcal{I}_{51}^{(4)}&=\text{cyc}\left[\frac{i x_{17}^4 x_{24}^2 \epsilon _{123458}}{x_{16}^2 x_{18}^2 x_{19}^2 x_{27}^2 x_{29}^2 x_{37}^2 x_{47}^2 x_{48}^2 x_{58}^2 x_{67}^2 x_{68}^2 x_{69}^2 x_{78}^2 x_{79}^2}\right] & \mathcal{I}_{52}^{(4)}&=\text{cyc}\left[\frac{i x_{13}^2 x_{14}^2 x_{37}^2 \epsilon _{123456}}{x_{16}^2 x_{17}^2 x_{19}^2 x_{29}^2 x_{36}^2 x_{38}^2 x_{39}^2 x_{47}^2 x_{48}^2 x_{57}^2 x_{67}^2 x_{68}^2 x_{69}^2 x_{78}^2}\right] \\
 \mathcal{I}_{53}^{(4)}&=\text{cyc}\left[\frac{i x_{14}^2 x_{17}^2 x_{28}^2 \epsilon _{123457}}{x_{16}^2 x_{18}^2 x_{19}^2 x_{27}^2 x_{29}^2 x_{37}^2 x_{47}^2 x_{48}^2 x_{58}^2 x_{67}^2 x_{68}^2 x_{69}^2 x_{78}^2 x_{79}^2}\right] & \mathcal{I}_{54}^{(4)}&=\text{cyc}\left[\frac{i x_{13}^2 x_{17}^2 x_{28}^2 \epsilon _{123458}}{x_{16}^2 x_{18}^2 x_{19}^2 x_{27}^2 x_{29}^2 x_{37}^2 x_{38}^2 x_{48}^2 x_{58}^2 x_{67}^2 x_{68}^2 x_{69}^2 x_{78}^2 x_{79}^2}\right] \\
 \mathcal{I}_{55}^{(4)}&=\text{cyc}\left[\frac{i x_{13}^2 x_{14}^2 x_{27}^2 \epsilon _{123457}}{x_{16}^2 x_{17}^2 x_{19}^2 x_{28}^2 x_{29}^2 x_{37}^2 x_{38}^2 x_{46}^2 x_{47}^2 x_{56}^2 x_{67}^2 x_{78}^2 x_{79}^2 x_{89}^2}\right] & \mathcal{I}_{56}^{(4)}&=\text{cyc}\left[\frac{i x_{13}^2 x_{46}^2 x_{56}^2 \epsilon _{123457}}{x_{16}^2 x_{19}^2 x_{26}^2 x_{36}^2 x_{38}^2 x_{47}^2 x_{48}^2 x_{57}^2 x_{59}^2 x_{67}^2 x_{68}^2 x_{69}^2 x_{78}^2 x_{79}^2}\right] \\
 \mathcal{I}_{57}^{(4)}&=\text{cyc}\left[\frac{i x_{17}^2 x_{24}^2 x_{36}^2 \epsilon _{123456}}{x_{16}^2 x_{18}^2 x_{27}^2 x_{28}^2 x_{37}^2 x_{39}^2 x_{46}^2 x_{49}^2 x_{56}^2 x_{67}^2 x_{68}^2 x_{69}^2 x_{78}^2 x_{79}^2}\right] & \mathcal{I}_{58}^{(4)}&=\text{cyc}\left[\frac{i x_{14}^2 x_{18}^2 x_{37}^2 \epsilon _{123456}}{x_{16}^2 x_{17}^2 x_{19}^2 x_{26}^2 x_{36}^2 x_{38}^2 x_{47}^2 x_{48}^2 x_{57}^2 x_{68}^2 x_{69}^2 x_{78}^2 x_{79}^2 x_{89}^2}\right] \\
 \mathcal{I}_{59}^{(4)}&=\text{cyc}\left[\frac{i x_{13}^2 x_{14}^2 x_{67}^2 \epsilon _{123459}}{x_{16}^2 x_{17}^2 x_{18}^2 x_{27}^2 x_{37}^2 x_{39}^2 x_{46}^2 x_{49}^2 x_{56}^2 x_{68}^2 x_{69}^2 x_{78}^2 x_{79}^2 x_{89}^2}\right] & \mathcal{I}_{60}^{(4)}&=\text{cyc}\left[\frac{i x_{13}^2 x_{26}^2 x_{47}^2 \epsilon _{123458}}{x_{17}^2 x_{18}^2 x_{27}^2 x_{29}^2 x_{36}^2 x_{39}^2 x_{46}^2 x_{48}^2 x_{58}^2 x_{67}^2 x_{68}^2 x_{69}^2 x_{78}^2 x_{79}^2}\right] \\
 \mathcal{I}_{61}^{(4)}&=\text{cyc}\left[\frac{i x_{16}^2 x_{24}^2 x_{57}^2 \epsilon _{123458}}{x_{17}^2 x_{19}^2 x_{26}^2 x_{27}^2 x_{36}^2 x_{46}^2 x_{48}^2 x_{58}^2 x_{59}^2 x_{67}^2 x_{68}^2 x_{78}^2 x_{79}^2 x_{89}^2}\right] & \mathcal{I}_{62}^{(4)}&=\text{cyc}\left[\frac{i x_{14}^4 x_{67}^2 \epsilon _{123456}}{x_{16}^2 x_{17}^2 x_{19}^2 x_{26}^2 x_{36}^2 x_{46}^2 x_{47}^2 x_{48}^2 x_{57}^2 x_{68}^2 x_{69}^2 x_{78}^2 x_{79}^2 x_{89}^2}\right] \\
 \mathcal{I}_{63}^{(4)}&=\text{cyc}\left[\frac{i x_{24}^2 \epsilon _{123456}}{x_{16}^2 x_{26}^2 x_{29}^2 x_{37}^2 x_{47}^2 x_{48}^2 x_{56}^2 x_{68}^2 x_{69}^2 x_{78}^2 x_{79}^2 x_{89}^2}\right] & \mathcal{I}_{64}^{(4)}&=\text{cyc}\left[\frac{i x_{13}^4 \epsilon _{123456}}{x_{16}^2 x_{17}^2 x_{19}^2 x_{29}^2 x_{36}^2 x_{38}^2 x_{39}^2 x_{48}^2 x_{57}^2 x_{67}^2 x_{68}^2 x_{69}^2 x_{78}^2}\right] \\
 \mathcal{I}_{65}^{(4)}&=\text{cyc}\left[\frac{i x_{13}^2 x_{14}^2 \epsilon _{123456}}{x_{16}^2 x_{17}^2 x_{19}^2 x_{29}^2 x_{38}^2 x_{39}^2 x_{46}^2 x_{48}^2 x_{56}^2 x_{67}^2 x_{68}^2 x_{78}^2 x_{79}^2}\right] & \mathcal{I}_{66}^{(4)}&=\text{cyc}\left[\frac{i x_{14}^2 x_{17}^2 \epsilon _{123456}}{x_{16}^2 x_{18}^2 x_{19}^2 x_{29}^2 x_{37}^2 x_{46}^2 x_{47}^2 x_{56}^2 x_{67}^2 x_{68}^2 x_{78}^2 x_{79}^2 x_{89}^2}\right] \\
 \mathcal{I}_{67}^{(4)}&=\text{cyc}\left[\frac{i x_{14}^2 x_{26}^2 \epsilon _{123456}}{x_{16}^2 x_{17}^2 x_{27}^2 x_{29}^2 x_{38}^2 x_{46}^2 x_{48}^2 x_{56}^2 x_{67}^2 x_{68}^2 x_{69}^2 x_{79}^2 x_{89}^2}\right] & \mathcal{I}_{68}^{(4)}&=\text{cyc}\left[\frac{i x_{13}^2 x_{26}^2 \epsilon _{123457}}{x_{17}^2 x_{18}^2 x_{27}^2 x_{29}^2 x_{36}^2 x_{39}^2 x_{46}^2 x_{58}^2 x_{67}^2 x_{68}^2 x_{69}^2 x_{78}^2 x_{79}^2}\right] \\
 \mathcal{I}_{69}^{(4)}&=\text{cyc}\left[\frac{i x_{16}^2 x_{24}^2 \epsilon _{123457}}{x_{17}^2 x_{18}^2 x_{26}^2 x_{27}^2 x_{36}^2 x_{46}^2 x_{49}^2 x_{58}^2 x_{67}^2 x_{69}^2 x_{78}^2 x_{79}^2 x_{89}^2}\right] & \mathcal{I}_{70}^{(4)}&=\text{cyc}\left[\frac{i x_{13}^2 x_{25}^2 \epsilon _{123456}}{x_{17}^2 x_{18}^2 x_{27}^2 x_{29}^2 x_{36}^2 x_{39}^2 x_{46}^2 x_{56}^2 x_{58}^2 x_{68}^2 x_{69}^2 x_{78}^2 x_{79}^2}\right] \\
 \mathcal{I}_{71}^{(4)}&=\text{cyc}\left[\frac{i x_{14}^2 x_{26}^2 x_{37}^2 \epsilon _{123458}}{x_{17}^2 x_{18}^2 x_{27}^2 x_{29}^2 x_{36}^2 x_{39}^2 x_{46}^2 x_{48}^2 x_{58}^2 x_{67}^2 x_{68}^2 x_{69}^2 x_{78}^2 x_{79}^2}\right] & \text{} \\